\documentclass[aps,pra,twocolumn,amsmath,amssymb]{revtex4-2}
\usepackage{adjustbox}
\usepackage[utf8]{inputenc}  
\usepackage{hyperref}        
\usepackage{booktabs}        
\usepackage{amsfonts}        
\usepackage{nicefrac}        
\usepackage{microtype}      
 
\usepackage{bm}
\usepackage{bbm}
\usepackage{braket}
\usepackage{graphicx}
\usepackage{subfig}
\usepackage{float}
\usepackage{amsthm}
\usepackage{algorithmic}
\usepackage[linesnumbered,ruled]{algorithm2e}
\SetKwInOut{Parameter}{parameter}

\usepackage{mathtools}
\usepackage{nccmath}
\usepackage{color}
\usepackage{caption}
\usepackage[dvipsnames]{xcolor}
\usepackage{titlesec}
\usepackage{ulem}
\usepackage{tikz}
\usetikzlibrary{quantikz}

\usepackage{soul}

\newcommand{\CZ}{\text{CZ}}

\newcommand{\X}{\text{X}}
\newcommand{\Y}{\text{Y}}
\newcommand{\Z}{\text{Z}}
\newcommand{\T}{\text{T}}

\newcommand{\R}{\text{R}}
\newcommand{\RY}{\text{R}_Y}
\newcommand{\RX}{\text{R}_X}
\newcommand{\RZZ}{\text{R}_{\text{ZZ}}}
\newcommand{\RZ}{\text{R}_{\text{Z}}}
\newcommand{\Tr}{\text{Tr}}

\newcommand{\agent}{\mathsf{A}}
\newcommand{\env}{\mathsf{E}}

\titlespacing{\title}{0pc}{0.1 pc}{0.3pc}
\titlespacing{\section}{0pc}{0.1pc}{0.3pc}

\begin{document}
	\title{Quantum Compiling with Reinforcement Learning on a Superconducting Processor}
 
	\author{Z. T. Wang$^{1,2,5}$}
    \thanks{These two authors contributed equally}
    \author{Qiuhao Chen$^{3,4}$}
	\thanks{These two authors contributed equally}
 
    \author{Yuxuan Du$^{6,4}$}
	\thanks{Corresponding authors}
	\email{duyuxuan123@gmail.com}

    \author{Z. H. Yang$^{2,5}$}

    \author{Xiaoxia Cai$^{1}$}

    \author{Kaixuan Huang$^{1}$}
    
    \author{Jingning Zhang$^{1}$}
    \author{Kai Xu$^{1,2,5,7}$}
    \author{Jun Du$^{1}$}
    \author{Yinan Li$^{3}$}
    \author{Yuling Jiao$^{3}$}
    
	\author{Xingyao Wu$^{4}$}
	\author{Wu Liu$^{4}$}
	\author{Xiliang Lu$^{3}$}
 \author{Huikai Xu$^{1}$}
 \author{Yirong Jin$^{1}$}
	\author{Ruixia Wang$^{1}$}
	\thanks{Corresponding authors}
	\email{wangrx@baqis.ac.cn}
		
	\author{Haifeng Yu$^{1,7}$}
	\thanks{Corresponding authors}
	\email{hfyu@baqis.ac.cn}
	
	\author{S. P. Zhao$^{1,2,5,7}$}
    
    \affiliation{$^1$Beijing Academy of Quantum Information Sciences, Beijing 100193, China}
	\affiliation{$^2$Beijing National Laboratory for Condensed Matter Physics, Institute of Physics, Chinese Academy of Sciences, Beijing 100190, China}
    \affiliation{$^3$School of Mathematics and Statistics, Wuhan University, Wuhan, China}
    \affiliation{$^4$JD Explore Academy, Beijing, 101111, China}
	\affiliation{$^5$School of Physical Sciences, University of Chinese Academy of Sciences, Beijing 100190, China}
 \affiliation{$^6$School of Computer Science and Engineering, Nanyang Technological University, Singapore 639798, Singapore}
    \affiliation{$^7$Hefei National Laboratory, Hefei 230088, China}

\begin{abstract}
	
To effectively implement quantum algorithms on noisy intermediate-scale quantum (NISQ) processors is a central task in modern quantum technology. NISQ processors feature tens to a few hundreds of noisy qubits with limited coherence times and gate operations with errors, so NISQ algorithms naturally require employing circuits of short lengths via quantum compilation. Here, we develop a reinforcement learning (RL)-based quantum compiler for a superconducting processor and demonstrate its capability of discovering novel and hardware-amenable circuits with short lengths. We show that for the three-qubit quantum Fourier transformation, a compiled circuit using only seven CZ gates with unity circuit fidelity can be achieved. The compiler is also able to find optimal circuits under device topological constraints, with lengths considerably shorter than those by the conventional method. Our study exemplifies the codesign of the software with hardware for efficient quantum compilation, offering valuable insights for the advancement of RL-based compilers.
\end{abstract}
	
	\maketitle
	
\section{Introduction}
	Quantum computers are expected to outperform classical ones in various applications~\cite{shor1999polynomial, daley2022practical, huang2022quantum, huang2022provably}. At present, the noisy intermediate-scale quantum (NISQ) processors already exist and will play an important role in the coming decades~\cite{preskill2018quantum,daley2022practical} before scalable fault-tolerant quantum processors are developed~\cite{preskill1998fault}. These NISQ processors consist of tens to a few hundred of noisy qubits with relatively short coherence time and gate operations with errors. Considerable effort has been devoted to maximizing their utility in different ways such as quantum circuit compilation~\cite{bharti2022noisy, ladd2010quantum, javadiabhari2014scaffcc, chong2017programming, paler2017fault, linke2017experimental, heim2020quantum, Jurcevic_2021} and quantum error mitigation~\cite{bharti2022noisy,endo2018practical,corcoles2019challenges,kusyk2021survey,cai2022quantum,huang2023near}. In particular, quantum compilation is the key step of quantum computation that translates quantum algorithms into hardware instructions for a quantum processor. For the NISQ processors, it is crucial to compile circuits with short length~\cite{bharti2022noisy} considering at the same time the qubit connectivity, gate expressivity, and codesign of the software and hardware~\cite{linke2017experimental, Jurcevic_2021} in order to achieve efficient quantum compilation.
	
\begin{figure*}[htbp]
		\centering
		\includegraphics[width=0.95\textwidth]{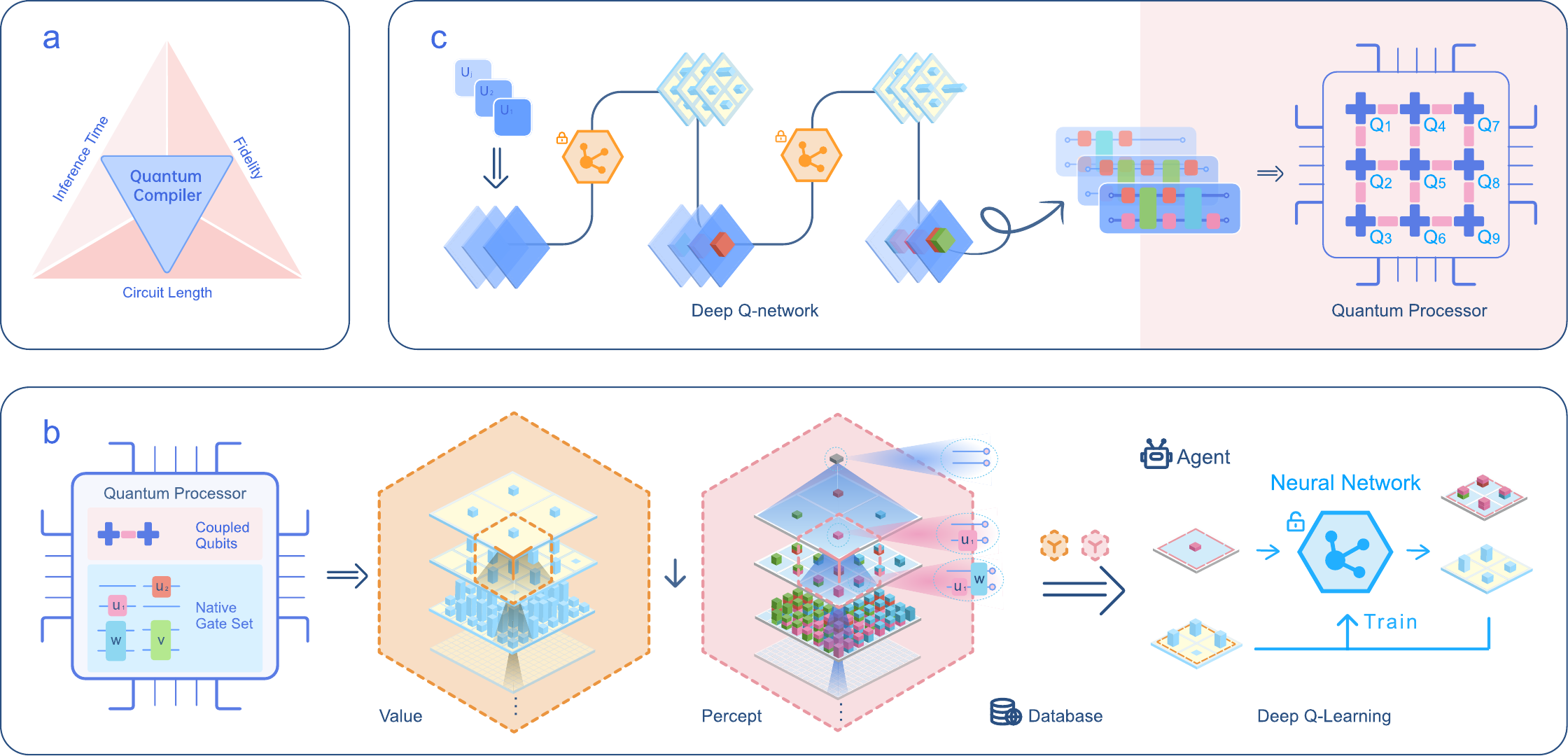}
		\caption{\label{fig: scheme}\small{\textbf{Architecture of the RL-based quantum compiler.} (a) Three metrics to measure the quality of a quantum compiler. (b) Precompilation. The action space is first initialized based on the native gate set and circuit topology of the superconducting processor. Then the agent starts the training of deep Q-network (DQN). Namely, DQN constantly queries the database, highlighted by two boxes `Percept' and `Value' in which small cuboids with different colors and heights represent different unitaries and values, to generate training data by pairing unitaries with the information of correct gate decompositions. The data training is from easy to hard along the downward arrow. (c) Inference. Given an unknown target unitary, the RL agent employs the AQ* search guided by the trained DQN to predict the decomposed gate sequence, which is finally implemented in the quantum processor. Here, compilation for the two-qubit system is illustrated for simplicity.}}
	\end{figure*}

	The performance of quantum compilers is in general characterized by three metrics: fidelity, circuit length, and inference time, as illustrated in Fig.~\ref{fig: scheme}(a). Since identifying the optimal sequence of quantum gates is NP-hard~\cite{botea2018complexity}, various quantum compilers have been developed and each class of quantum compilers holds its own strengths and weaknesses. For instance, Solovay-Kitaev based compilers~\cite{kitaev1997quantum, dawson2005solovay, kitaev2002classical} excel in compiling accuracy with the ability to achieve near-perfect fidelity, but at the cost of longer inference time and circuit length, and with the limitation of compiling gates for at most two qubits. To address these issues, heuristic compilers have been proposed~\cite{davis2019heuristics, younis2020qfast, khatri2019quantum, sharma2020noise, xu2021variational, madden2022best, rakyta2022approaching, jones2022robust, herrera2022policy}, which use greedy search and variational strategy to estimate the optimal gate sequence, with the central goal to reduce the gate count and circuit length in quantum applications.
	
Reinforcement learning (RL), as a general way of optimizing strategies with complex decision-making problems,  has recently emerged as a promising approach to address limitations in quantum compilers. This approach includes a precompilation procedure enabling manifest reduction of the overall execution time~\cite{moro2021quantum}. Numerical studies have verified their abilities in generating near-optimal gate sequences with high accuracy in single- and two-qubit systems~\cite{zhang2020topological, chen2022efficient}. In addition, a quantum compiler for an eight-qubit system is proposed by training RL models to predict parameterized compiling circuit structures with a variational quantum algorithm~\cite{he2021variational}. RL algorithms are also tailored for qubit routing to minimize SWAP gate for real-world hardware execution~\cite{herbert2019UsingReinforcementLearning, pozzi2020using}. Despite the progress, RL-based compilations on real quantum processors and their performance remain largely unexplored.
	
In this work, we develop a quantum compiler based on deep Q-network (DQN) and AQ* search strategy~\cite{chen2022efficient} for a superconducting multiqubit processor with a given qubit connectivity and native gate set. Conceptually, the DQN alleviates the challenge of exponentially increasing search space, and the AQ* search provides an efficient inference. The synergy between DQN and AQ* search guarantees efficiency in multiqubit compiling tasks. We conduct extensive experiments on the processor to show that the compiler is able to discover novel and hardware-amenable circuits with short lengths, showcasing remarkable performance in two- and three-qubit systems. Particularly, we demonstrate that for the three-qubit quantum Fourier transformation (QFT), the compiler can find circuits having only seven CZ gates with unity circuit fidelity. In the presence of hardware topological constraints, it is able to compile circuits with lengths considerably shorter than those by conventional methods. Our work offers an example of codesigning the software with hardware in quantum compilation and allows for a practical assessment of quantum compilers for the NISQ processors.
	
	\begin{figure*}[htbp]
		\centering
		\includegraphics[width=0.9\textwidth]{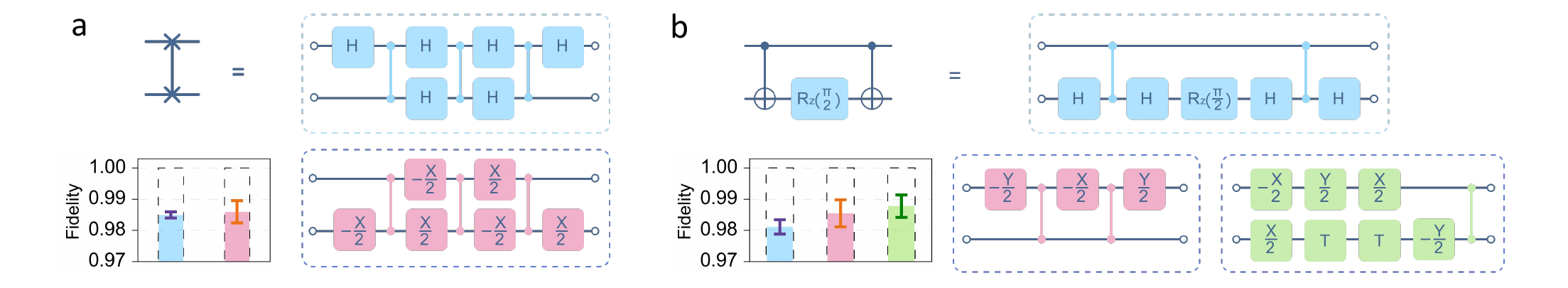}
		\caption{\small{\textbf{Compilation for two-qubit system.} (a) SWAP gate. The blue-colored circuit represents the standard decomposition, while the pink-colored circuit shows the result by RL compilation with the inference time of $\sim 1.23$ seconds. The lower-left panel displays the statistical results of fidelity $F_2$ measured on Q$_1$-Q$_2$ for each circuit with the same color, and histograms drawn with dashed lines refer to the results of $F_1$. (b) Similar results for $\RZZ(\pi/{2})$ gate with pink and green colors representing two gate sequences returned by the RL compiler. The inference times are comparable to that for the SWAP gate. 
		}}
		\label{fig:exp_two_qubits}
	\end{figure*}
	
	\medskip
	\section{Results}
	
	\subsection{RL-based compiler and experimental system}
	
	RL-based compiling starts with the considerations of the native gate set $\mathcal{G}$ and the device topology of a quantum processor. In our experiment, we use a $9$-qubit superconducting processor with tunable qubit coupling and investigate the performance of the circuits from the RL-based compiler (or RL compiler) for the two- and three-qubit chains formed in the processor. A native gate set $\mathcal{G} = \{ \pm \X/2,\pm \Y/2, \T, \T^{\dagger}, \CZ\}$ is used and the average single- and two-qubit gate fidelities are about $99.9\%$  and $99.3\%$, respectively (see Supplementary Information~A for details). For a given target unitary $\hat{U}$, the compiler returns, within the reference time, a gate sequence $\{\tilde{G}_i\}_{i=1}^{N}\in \tilde{\mathcal{G}}^N$ approximating $\hat{U}$, where $\tilde{\mathcal{G}}$ is the generalized gate set considering device topology. For example, in the two-qubit case, the $\T$ gate in $\mathcal{G}$ is extended to $\T\otimes \mathbb{I},  \mathbb{I}\otimes \T \in  \tilde{\mathcal{G}}$. We follow conventions to measure the performance of the RL compiler in terms of the gate sequence length (or circuit length), the circuit fidelity, and the inference time. For the circuit length, it amounts to the total gate number $N=N_1+N_2$ in $\{\tilde{G}_i\}_{i=1}^{N}$, where $N_1$ and  $N_2$ denote the number of single- and two-qubit gates, respectively. The circuit fidelity is evaluated by $F_1(\hat{U},\hat{V})=\Tr(\hat{U}^{\dagger}\hat{V})/2^M$, where $M$ is the number of qubits and $\hat{V}=\prod_{i=1}^{N} \tilde{G}_i$ is the circuit unitary~\cite{Bao2022FluxoniumAA}. The corresponding experimental fidelity is obtained by quantum process tomography (QPT), i.e.,  $F_2(\hat{U}, \hat{V}) = \Tr ( \mathcal{\hat{E}}^{\dagger} \hat{\mathcal{F}})$, where $\mathcal{\hat{E}}$ and $\hat{\mathcal{F}}$ are the calculated and measured process matrices of $\hat{U}$ and $\hat{V}$, respectively. 
	
	The compiling process has two stages: precompilation and inference. Precompilation aims to train a learning model that decomposes any unitary $\hat{U}$ into an optimal sequence of gates in $\tilde{\mathcal{G}}$~\cite{zhang2020topological} and can be approached by deep Q-learning algorithm to capture the target learning model, dubbed deep Q-Network (DQN)~\cite{sutton2018reinforcement}. As shown in Fig.~\ref{fig: scheme}(b), the action space $\mathcal{A}=\{A_i\}_{i=1}^d$, equivalent to $\tilde{\mathcal{G}}$, contains elements representing legitimate operations on the quantum processor. Once the action space is settled, the environment $\env$ (a.k.a, database) and the agent $\agent$ interact to learn collaboratively with in total $L$ loops. Specifically, the first loop with $l=1$ starts from the identity operation $\mathbb{I}$, as $\env$ applies each action in $\mathcal{A}$ to $\mathbb{I}$ to synthesize $d$ percepts (unitaries) $\mathcal{W}_1=\{W_k\}_{k=1}^d$ associated with $d$ values $\mathcal{V}_1=\{V_k\}_{k=1}^d$. These values represent the minimum number of gates required to revert the corresponding percept back to $\mathbb{I}$ with a minus sign, i.e, $W_k = A_k$ and $V_k=-1$ for $\forall k\in[d]$. Following the same routine in the second loop with $l=2$, $\env$  applies each action in $\mathcal{A}$ to $\mathcal{W}_1$ to synthesize $\mathcal{W}_2$ and calculates the corresponding values $\mathcal{V}_2$. The cardinality of $\mathcal{W}_2$ is $d^2$ and there are two possible outcomes of $0$ or $-2$. The prepared $\mathcal{W}_1$, $\mathcal{A}$, and $\mathcal{V}_2$ are sent to $\agent$ to conduct learning, which is accomplished by the DQN. The Q-value function is utilized to guide the optimization of DQN to minimize the distance between the predictions and the ground truth~\cite{sutton2018reinforcement}.  After training $T$ epochs, $\agent$ sends a terminal feedback to $\env$, which completes the current loop. In the next loop $l=3$, $\env$ synthesizes more complex percepts $\mathcal{W}_3$ and calculates the corresponding values $\mathcal{V}_3$, and feeds $(\mathcal{W}_2, \mathcal{A}, \mathcal{V}_3)$ to $\agent$ to train DQN. Progressively, the percepts $\mathcal{W}_l$ become increasingly complicated and gradually cover the whole unitary space when the total loop number $L\rightarrow \infty$,   implying that DQN is forced to learn a general rule for gate decomposition (see Methods).
	
	The inference, as depicted in Fig.~\ref{fig: scheme}(c), leverages the trained DQN to decompose any unitary $\hat{U}$ into a sequence of native gates. This process utilizes the AQ* search algorithm~\cite{agostinelli2021search}, which iteratively expands the percept with the highest value until the target identity is reached or a pre-defined search depth is exceeded. Below we will see that circuits with the shortest length to date can be obtained from the AQ* search, and the circuit fidelity $F_1$ further increases after optimizing single-qubit gate parameters using a post-processing toolkit based on the variational principle~\cite{mclachlan1964variational}. We will dub our compiler with this post-processing as a variational RL-based (VRL) compiler (see Methods for details). Note that for superconducting processors, the parametrization of single-qubit gates does not increase any gate count and gate operation time, and can be conveniently implemented~\cite{PhysRevA.96.022330}. 
	\begin{figure*}[htbp]
		\centering
		\includegraphics[width=1.0\textwidth]{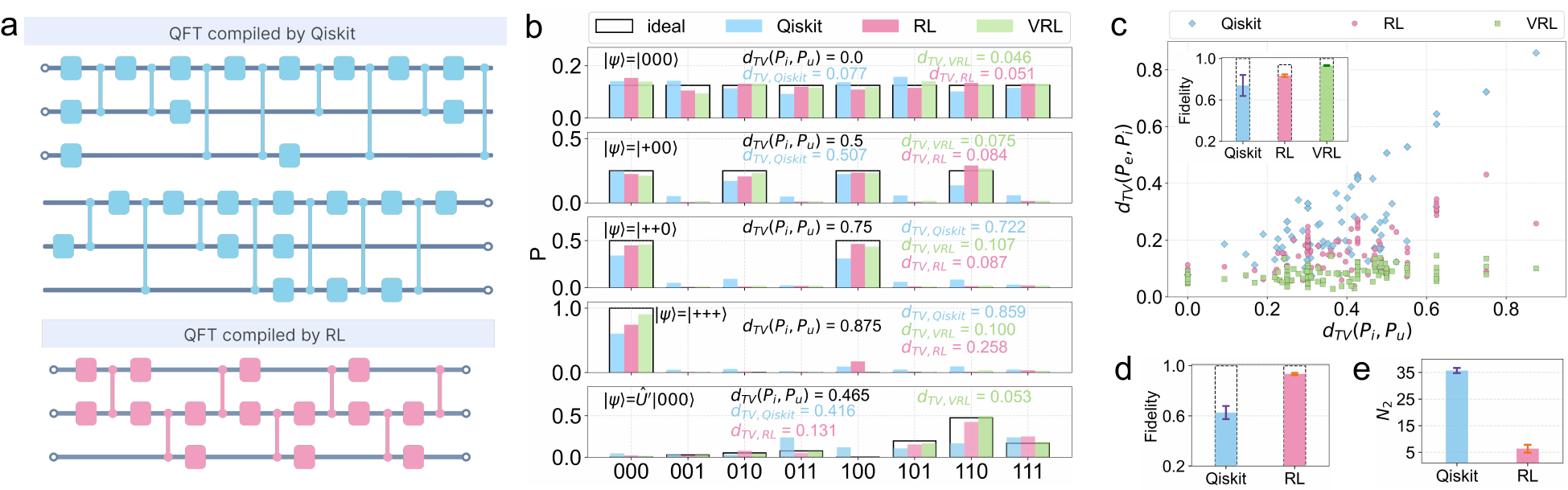}
		\caption{\small{\textbf{Compilation for the three-qubit system.} (a) QFT circuits by Qiskit and RL compiler (inference time $\sim$ 800 seconds). Squares denote single-qubit gates from $\mathcal{G}$. (b) Performance analysis of the QFT circuits on Q$_2$-Q$_3$-Q$_6$ qubits using the total variance (TV) metric. The QFT circuit is evaluated on four varied input states $|000\rangle$, $|$+$00\rangle$, $|$++$0\rangle$, $|$+++$\rangle$ and $\hat{U}'|000\rangle$, where $\hat{U}'$ is randomly sampled from SU(8). Here, $d_{\text{TV}}(P_i, P_u)$ quantifies the TV distance between the distributions of the ideal output state $P_i$ and the uniform distribution $P_u$ in the  measurement of  computational basis. For each panel in Subplot~(b), the notations $d_{\text{TV,Qiskit}}$, $d_{\text{TV,RL}}$, and $d_{\text{TV,VRL}}$ represent $d_{\text{TV}}(P_e, P_u)$ in which $P_e$ refers to the distribution of the generated state for QFT circuits compiled by Qiskit, RL, and VRL compilers, respectively. A lower difference with  $d_{\text{TV}}(P_i, P_u)$ suggests a better performance. (c) Statistical performance comparison of Qiskit, RL, and VRL by preparing arbitrary input states. The notaions in x-axis and y-axis follow those in Subplot~(b). The inset shows the fidelity (dashed line refers to $F_1$ and histogram refers to $F_2$) of the QFT circuits compiled by three indicated compilers. Subplots (d) and (e) analyze the statistical performance of Qiskit and RL compilers for compiling $10$ two-qubit gates with topological constraints under the metric of $F_1$ (dashed line), $F_2$ and $N_2$.}}
		\label{fig:three_qubit_experiment}
	\end{figure*}

	\subsection{Compilation for two-qubit system}
	
	To compile the two-qubit gates, we build DQN through an adaptation of ResNet architecture~\cite{he2016deep} with 60 million parameters, and the loop number is set as $45$. The training dataset comprises 3.02 billion examples, as detailed in Methods.
	
    Figure~\ref{fig:exp_two_qubits} shows the compiling and experimental results of the widely used SWAP and $\RZZ$ gates. The experiment is performed on Q$_1$-Q$_2$ (see chip schematic on the right side in Fig.~\ref{fig: scheme}(c)) and each decomposed gate sequence is performed five times repeatedly to evaluate the statistical performance. For the SWAP gate, the gate sequences by the standard decomposition and RL compiler show optimal results, both with three CZ gates ($N_2=3$) and circuit fidelity $F_1$=1. However, the replacement of the H gate by the $\pm$X/2 gates enables the RL compiler to achieve a higher experimental fidelity $F_2=0.9859\pm 0.0035$ compared to $F_2=0.9849 \pm 0.0010$ from the standard compilation. This improvement arises because implementing an H gate typically requires multiple single-qubit gates from $\mathcal{G}$.
    With $\RZZ(\pi/2)$, our compiler returns two solutions. The first solution has the same CZ gate number ($N_2=2$) as the standard solution but differs in the adopted single-qubit gates. As with the SWAP gate, this difference leads to a higher fidelity $F_2=0.9854\pm 0.0043$ compared to $F_2=0.9811\pm 0.0023$ in the standard case. With the second solution, only one CZ gate ($N_2=1$) is used and a fidelity $F_2=0.9877 \pm 0.0036$ is achieved. These results highlight the capability of the RL compiler to discover novel circuits with a reduced number of two-qubit gates. Results for more general two-qubit targets, including random gates from SU(4), are discussed in Supplementary Information~C.
	
	\subsection{Compilation for three-qubit system}
	
	For the compilation of three-qubit systems, the loop number in training DQN is set to $L=44$, resulting in a training dataset of 27.4 billion examples. The DQN architecture is identical to the two-qubit case, but with an increased parameter space to enhance performance during inference.  To demonstrate the performance of the compiler, we first consider QFT, a key operation in many quantum algorithms to gain speedups. The constraint of $\CZ$ gate applied only to the nearest-neighbor qubits in the chain is considered. In Fig.~\ref{fig:three_qubit_experiment}(a), we find that the RL compiler yields a circuit with CZ gate count of $N_2=7$, as compared to the Qiskit compiler~\cite{qiskit} with $N_2 = 15$ and compilers employing heuristic search~\cite{davis2019heuristics,PhysRevX.11.041039,lu2023QuantumCompilingVariational} with lowest $N_2=8$. 
	
The distributions of the output states by QFT for the initial states $|000\rangle$, $|$+$00\rangle$, $|$++$0\rangle$, $|$+++$\rangle$, and a random state are visualized in Fig.~\ref{fig:three_qubit_experiment}(b), where the ideal results and the experimental data measured on Q$_2$-Q$_3$-Q$_6$ for the circuits by the Qiskit, RL, and VRL compilers, are presented. We observe successive improvements in the experimental results for Qiskit, RL, and VRL compilers. This can be seen more clearly from the total variation (TV) distance (detailed in Supplementary Information~B) introduced to measure the distance between the experimental and ideal output distributions~\cite{PhysRevX.11.041039}. We prepare stochastic initial states and the experimental results of TV distance are shown in Fig.~\ref{fig:three_qubit_experiment}(c). In the inset of Fig.~\ref{fig:three_qubit_experiment}(c), the experimental fidelities $F_2$ between the target unitary and the circuits returned by the Qiskit, RL, and VRL compliers are presented, showing the same trend of improvements. From the TV distance and the fidelity, the experimental results from the RL-compiled circuit are still better than those from the circuit by Qiskit, even though they have the fidelities $F_1$ of 0.94 and 1, respectively. These results unveil the interplay among circuit fidelity, circuit length and structure, and gate errors and decoherence of the quantum processor.
	
We next consider the compilation of two-qubit gates between $\text{Q}_2$ and $\text{Q}_6$ in the Q$_2$-Q$_3$-Q$_6$ chain, which may occur for devices with limited qubit connectivity and can only be realized via CZ gates between $\text{Q}_2$ and $\text{Q}_3$ or $\text{Q}_3$ and $\text{Q}_6$. In Figs.~\ref{fig:three_qubit_experiment}(d) and (e), we show the results of compiling $10$ random two-qubit gates. On average, the RL compiler uses $N_2=6.36$ $\CZ$ gates with fidelity $F_2=0.9342\pm 0.0089$, whereas the Qiskit compiler uses $N_2=35.72$ $\CZ$ gates with fidelity $F_2=0.6264\pm 0.0525$. These results show a clear advantage of the RL compiler in the presence of device topological constraints. The advantage can be attributed in part to the compiling approach of starting from the matrix representation of the target operator and globally seeking the gate sequence, whereas conventional compilers typically begin with an initial circuit guess and locally optimize the structure or parameters.
	
	\subsection{Experimental error analysis}
	
\begin{figure}[t] 
		\centering
		\includegraphics[width=0.48\textwidth]{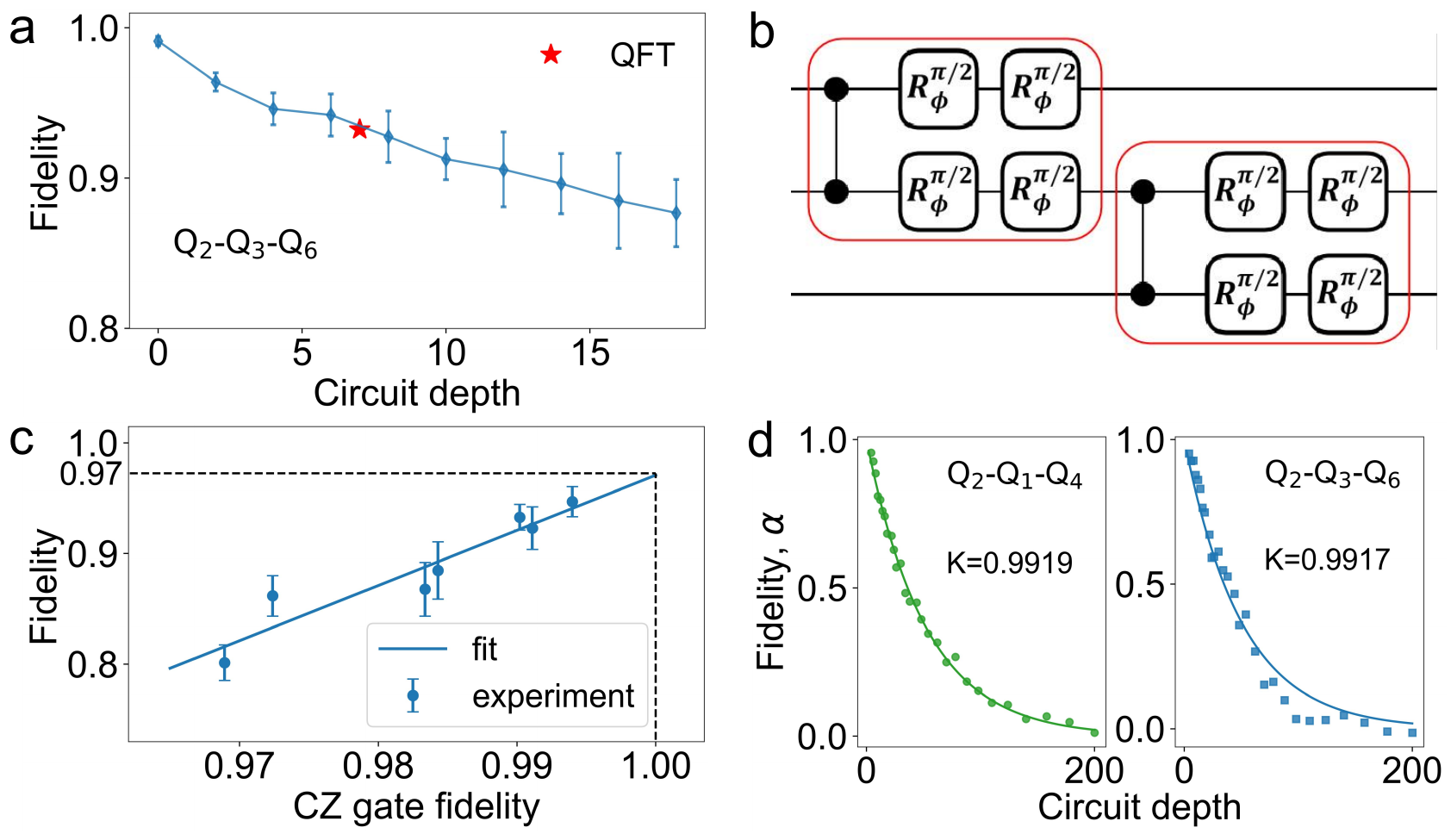}
		\caption{\small{\textbf{Error analysis.} (a) Experimental fidelity measured by QPT versus the unit circuit depth with random circuits shown in (b) for Q$_2$-Q$_3$-Q$_6$. At the zeroth depth, only two $\text{R}_{\phi}(\pi/2)$ gates are applied to each qubit, where  $\text{R}_{\phi}(\pi/2)=\exp({-i\frac{\pi}{4}(\cos{\phi}\text{X}+\sin{\phi}\text{Y})})$ and $\phi$ refers to a set of specific angles. The star indicates the circuit in Fig.~\ref{fig:three_qubit_experiment}(a) by RL compiler. (c) Experimental fidelity of QFT circuit measured by QPT versus averaged CZ gate fidelity for seven three-qubit sets of Q$_2$-Q$_1$-Q$_4$, Q$_2$-Q$_3$-Q$_6$, Q$_3$-Q$_6$-Q$_5$, Q$_4$-Q$_5$-Q$_2$, Q$_4$-Q$_5$-Q$_6$, Q$_5$-Q$_2$-Q$_3$, and Q$_6$-Q$_5$-Q$_2$. The line is a linear fit. (d) Circuit depth dependences of fidelity measured using XEB for two three-qubit sets. The lines fit as discussed in the text.}}
		\label{fig:fig4}
	\end{figure}

	For NISQ processors, the experimental fidelity $F_2$ is always lower than the circuit fidelity $F_1$ due to the imperfect gate operation, qubit decoherence, and initial-state preparation and measurement errors. In Fig.~\ref{fig:fig4}(a), we present the results of $F_2$ as a function of circuit depth, showing a clear decrease in fidelity with increasing depth. As shown in Fig.~\ref{fig:fig4}(b), here we take one block, each containing a CZ gate and 4 random $\text{R}_{\phi}(\pi/2)$ gates, as the unit circuit depth metric and apply it to the upper and lower two qubits alternately in the three-qubit chain. Besides, for the zero circuit depth, it refers to two $\text{R}_{\phi}(\pi/2)$ gates applied to each qubit, where experimental results indicate its fidelity ranging from 0.991 to 0.994.  This means that the errors from initial-state preparation and final-state readout are small compared to the gate errors from the six single-qubit gates with an average fidelity of 0.999. Fig.~\ref{fig:fig4}(c) shows $F_2$ versus CZ gate fidelity. Each CZ gate fidelity is an average over all CZ gates in the QFT circuit for seven three-qubit sets with intentionally reduced gate fidelities (see also Supplementary Information~A). As indicated by the dashed lines, extrapolating the CZ gate fidelity to unity would result in a fidelity of about 0.97, where the discrepancy to the perfect fidelity is caused primarily by the imperfect operations of 28 single-qubit gates in the circuit. These results indicate that for shallow circuits (with short operation time relative to qubit coherence time), the gate fidelity is the dominant factor for the experimental performance.
	
We use cross entropy benchmarking (XEB)~\cite{CharacterizingQuantumSupremacy,doi:10.1126/science.aao4309}(see Supplementary Information~A) to measure fidelity as a function of circuit depth, which allows for the measurement with much deeper depth. The results for two three-qubit sets are shown in Fig.~\ref{fig:fig4}(d) with the depth extending to 200. 	
The data align well with the heuristic fitting function~\cite{kharkov2022ArlineBenchmarksAutomated, Jurcevic_2021}: $\alpha(d)$ = $K^d\cdot F_{\rm CZU}^{m} \cdot F_{\rm CZL}^{d-m}\cdot F_{\text{R}_{\phi}(\pi/2)}^{4d}$. Here $F_{\rm CZU}$ and $F_{\rm CZL}$ denote the CZ fidelities for the upper and lower qubit pairs, $F_{\text{R}_{\phi}(\pi/2)}$ is the average single-qubit gate fidelity, $d$ is the circuit depth, and $m$ is the number of blocks applied on the upper two qubits. A fitting parameter $K \approx$ 0.992 is obtained, which accounts for errors uncovered by gate infidelities such as decoherence, non-Markovian noise, and crosstalk, etc. A $K$ value comparable to the CZ gate fidelity implies a comparable role played by factors other than gate fidelity for deep circuits.

	\medskip
	\section{Discussion and outlook}
	
	We have developed an RL-based compiler for the superconducting processor with given qubit connectivity, native gate set, gate fidelity, and coherence properties. The compiler shows excellent performance in finding short circuits with high fidelity and in experimental implementations. The influence of decoherence as well as gate and measurement errors is experimentally investigated, which shows the requirement of better device properties for further improvements and may help predict the performance of RL-based compilers on a specific superconducting processor. Compared to the previous RL-based compilers~\cite{zhang2020topological, moro2021quantum}, our DQN model leverages more information about percept performance, leading to efficient training and inference, thereby facilitating effective compilation for multiqubit targets. A future direction will be to design more informative RL models~\cite{fawziDiscoveringFasterMatrix2022, jumper2021HighlyAccurateProtein} for training, particularly for scenarios of increasing qubit number. Incorporating supervised learning techniques during training can penalize the RL model for generating redundant gate sequences~\cite{NIPS2017_d5e2c0ad, ouyang2022TrainingLanguageModels}, thus promoting efficient inference. Furthermore, exploring the potential of RL-based compilers for simultaneous circuit optimization~\cite{fosel2021quantum, ostaszewski2021reinforcement, dueck2018optimization} and gate decomposition~\cite{vartiainen2004efficient, paige1994history, krol2022efficient, mottonen2004quantum, kharkov2022ArlineBenchmarksAutomated} holds promise for significantly improving compilation efficiency.
		
	\medskip
	\section{Methods}{\label{methods}}
	 Here we introduce the technical details in the precompilation stage, including training data generation and the optimization of DQN. We also describe the implementation of AQ* search, exploited in the inference stage. 
	
	\subsection{Training data generation} For training data generation, we adopt two techniques that are widely used in reinforcement learning to improve the performance and robustness of a compiler. 
	
	The first technique is adding a bounded perturbation to the initial operation $\mathbb{I}$ ~\cite{zhang2020topological}. The perturbed operation is denoted by $\mathbb{I}'$ satisfying $F_1(\mathbb{I}, \mathbb{I}')< 10^{-4}$. The employment of a set of $\{\mathbb{I}'\}$  effectively mitigates the limitation of merely using $\mathbb{I}$ in the sense that all expansions stem from a single percept. 
	
	The second technique is injecting randomness in preparing $\mathcal{W}_l$ and $\mathcal{V}_l$ when $l$ becomes large. Note that the construction rule of $\mathcal{W}_l$ and $\mathcal{V}_l$ presented in the main text implies that their cardinalities exponentially scale with $l$. Such computational overhead and the memory cost become prohibitive for large $l$. For this reason, when $l$ exceeds a threshold, we only collect a fraction of all possible synthesized unitaries to form $\mathcal{W}_l$. The quantity of this threshold depends on the problem in hand and the accessible memories of classical hardware, as detailed in Supplementary Information~B. 
	
	\subsection{Optimization of DQN} As discussed in the main text, when $l=2$, the prepared $(\mathcal{W}_1, \mathcal{A}, \mathcal{V}_2)$ is sent to the agent $\agent$ to conduct learning. This process is completed by DQN guided by the Q-value function. Here we first expand this process and then extend the optimization of DQN to the general setting. 
	
	When $l=2$, the percepts $\mathcal{W}_2$ received by $\agent$ take the form $\mathcal{W}_2=\{A_1A_1,\cdots,A_kA_{k'},\cdots,A_dA_d\}$ and the values in $\mathcal{V}_2$ contain two outcomes, i.e.,  $V_{kk'}\in\{0, -2\}$ for $\forall k,k'\in[d]$. As shown in Fig.~\ref{fig: scheme}(b), given these training data,  DQN $Q_{\bm{\theta}}$ takes $W_i\in \mathcal{W}_1$ as input and predicts the values of $[A_1W_i, A_2W_i,...,A_dW_i]$ as $\hat{\bm{y}}= [Q_{\bm{\theta}}(W_i,A_1), Q_{\bm{\theta}}(W_i,A_2),...,Q_{\bm{\theta}}(W_i,A_d)]\in \mathbb{R}^{d}$, where the trainable parameters are denoted by $\bm{\theta}$. The optimization of $\bm{\theta}$, which aims to minimize the distance between the predictions $\hat{\bm{y}}$ and the ground truth $\bm{y}=[V_{1i}, V_{2i},\cdots, V_{di}]\in \{-2,0\}^d$, is guided by the Q-value function in the mathematical form  
	\begin{equation}\label{eq:MES_loss}
		\mathcal{L}_{l}(\bm{\theta}) = \frac{1}{d\cdot|\mathcal{W}_l|} \sum_{W_i\in \mathcal{W}_l}\sum_{j=1}^d \left(Q_{\bm{\theta}}(W_i, A_j)-V_{ji} \right)^2,
	\end{equation}
	with setting $l=2$. The gradient descent optimizer is employed to update $\bm{\theta}$ to minimize $ \mathcal{L}_{l}(\bm{\theta})$ with $T$ epochs. This completes the optimization with $l=2$. 
	
	Following the same routine, at the $l$-th loop, we feed $(\mathcal{W}_{l-1}, \mathcal{A}, \mathcal{V}_l)$ to $\agent$ to conduct learning. More specifically, the gradient descent optimizer is employed to update $\bm{\theta}$ to minimize $\mathcal{L}_{l}(\bm{\theta})$. The optimization of the $l$-th loop is completed after $T$ epochs.
	
	We apply a slight modification to the loss $\mathcal{L}_{l}(\bm{\theta})$ when $l$ exceeds a threshold value. Such a modification originates from the fact that for large $l$, the acquisition of the values $\mathcal{V}_l$ is computationally expensive, which prohibits the optimization of $\mathcal{L}_{l}(\bm{\theta})$. To this end, we replace the values $\mathcal{V}_l$ by its estimation $\hat{\mathcal{V}}_l$, i.e., 
     \begin{eqnarray}
         \hat{V}_{ji} = \max_{A\in\mathcal{A}} \left( Q_{\bm{\theta}}(A_jW_i, A)-1 \right),~\forall j. \nonumber
     \end{eqnarray}
     Accordingly, the loss function is rewritten as 
	\begin{equation}\label{eq:Bellman_operator}
		\mathcal{L}_{l}(\bm{\theta}) = \frac{1}{d\cdot|\mathcal{W}_l|} \sum_{W_i\in \mathcal{W}_l}\sum_{j=1}^d \left(Q_{\bm{\theta}}(W_i, A_j)-\hat{V}_{ji} \right)^2.
	\end{equation}
	Supported by the theoretical analysis in ~\cite{bertsekas2019reinforcement,sutton2018reinforcement}, the optimization of the above two equations is equivalent when $L \rightarrow \infty$. The algorithmic details of DQN are deferred to Supplementary Information~B.
	
	\subsection{AQ* Search}
	 AQ* search is a heuristic-based search algorithm designed to find a path between a starting node $p_s$ and a goal node $p_t$ on a graph structure ~\cite{agostinelli2021search}. In quantum compiling, each node corresponds to a percept representing a synthesized unitary, the starting node and the goal node correspond to the target unitary $\hat{U}$ and the initial percept $\mathbb{I}$ (or $\mathbb{I}'$), respectively. As indicated by ~\cite{agostinelli2019solving,agostinelli2021search}, A* search ~\cite{zeng2009finding} or AQ* search can solve Rubik’s cube, or equivalently quantum compiling, using shorter solutions in less time than Monte Carlo tree search. 
	
	We recap the workflow of AQ* search for self-consistency~\cite{agostinelli2021search}. Concisely, we denote a node by $p$, which stands for a percept (or equivalently a synthesized unitary) in $\{\mathcal{W}_l\}_{l=1}^L$.  AQ* search leverages an evaluation function 
    \begin{eqnarray}
        f(p, A; \hat{U})=g(p ; \hat{U})+Q_{\bm{\theta}}(p, A) \nonumber
    \end{eqnarray}
    to find the path between $p_s$ to $p_t$, where $A\in \mathcal{A}$ refers to an action, $\hat{U}$ represents the target unitary, $g(p; \hat{U})$ counts the number of the exploited actions from $\hat{U}$ to $p$ during the search with a minus sign, and $Q_{\bm{\theta}}(p, a)$ is the output of the trained DQN after optimizing Eq.~(\ref{eq:MES_loss}). Supported by the evaluation function $f(\cdot,\cdot)$, AQ* search seeks the solution in an iterative manner. It maintains a set of intermediate percepts $\{\hat{p}\}$, which initially only contains the starting node $p_i$. In each iteration, for all $p'\in \{\hat{p}\}$, the algorithm expands $p'$ with a successor node $p^{\star}=A^{\star}p'$  by leveraging the evaluation function with 
    \begin{eqnarray}
        A^{\star}=\arg \max_{A \in \mathcal{A}} f(p', A ; \hat{U}). \nonumber
    \end{eqnarray}
    All the successor nodes are added into $\{\hat{p}\}$ if they have not been encountered before. Note that this process is efficient since the expansion for different $p'$ can be executed in parallel. The iteration process is terminated depending on two conditions. First, a node $p'\in \{\hat{p}\}$ has approached the goal node $p_t$ within a small error, i.e.,  $1-F_1(p', p_t)<10^{-4}$. Second, the runtime of searching exceeds a predefined threshold. In this case, AQ* search outputs a node in $\{\hat{p}\}$ closest to $p_t$.
	
	\subsection{Variational RL-based Compiler} 
	The variational RL-based compiler adheres to the variational principle by optimizing parameterized quantum gates while simulating the quantum operations classically. Similar ideals have also been broadly used in tensor networks~\cite{Cirac2021Matrix}. In particular, given a target operation $\hat{U}$, the nodes (percepts) set $\{\hat{p}\}$ generated by the AQ* Search can be transformed into a sequence of quantum gates $\{\tilde{G}_i\in\tilde{\mathcal{G}}\}$ satisfying $\hat{V}=\prod \tilde{G}_i\approx \hat{U}$. Notably, the gate set $\tilde{\mathcal{G}}$ is pre-fixed during compiling and remains fixed throughout the process, limiting potential optimization opportunities. The proposed variational RL-based compiler operates by taking $\hat{U}$ and $\hat{V}$ as inputs. It then  parameterizes all single-qubit gates within $\{\tilde{G}_i\}$ and optimizes the resulting variational ansatz $\hat{V}_{\boldsymbol{\gamma}}=\prod \tilde{G}_i(\gamma_i)$ towards $\hat{U}$ on a classical computer. The loss function is defined as \[\mathcal{L}(\hat{V}_{\bm{\gamma}}, \hat{U}) = \Tr\left( \hat{U}^{\dagger} \cdot \hat{V}_{\bm{\gamma}} \right). \]
	Subsequently, a gradient descent algorithm is used to adjust the parameter vector \(\bm{\gamma}\) and minimize this loss function. 
	
 \section*{Acknowledgements}  
 We thank H. L. Sheng for help in the preparation of figures. Numerical calculations were performed on the supercomputing system of the Supercomputing Center at Wuhan University. This work was supported by the National Natural Science Foundation of China (Grants Nos. 92365206, 21973003, 22288201, 22303005, 12104056, and T2322030), the Fundamental Research Funds for the Central Universities and Innovation Program for Quantum Science and Technology (Grant No. 2021ZD0301800), the National Nature Science Foundation of Hubei (Grant No. 2024AFA045) and the Beijing Nova Program (Grant No. 2022000216).

  \newpage
  \clearpage
   
	\onecolumngrid
	
	\renewcommand{\thefigure}{S\arabic{figure}}	
        \renewcommand{\thetable}{S\Roman{table}}
	\setcounter{figure}{0}

 \appendix
 
\renewcommand{\thesection}{SI~\Alph{section}}
\renewcommand{\thesubsection}{SI~\Alph{section}.\arabic{subsection}}

\begin{center}
	{\Large Supplementary Information of ``Quantum Compiling with Reinforcement Learning on a Superconducting Processor"}
\end{center}

	\tableofcontents

\vspace{10mm}
\section{EXPERIMENTAL DETAILS}\label{append:exp_setup}
\vspace{5mm}

\begin{figure*}[t]
	\centering
	\includegraphics[width=1.0\textwidth]{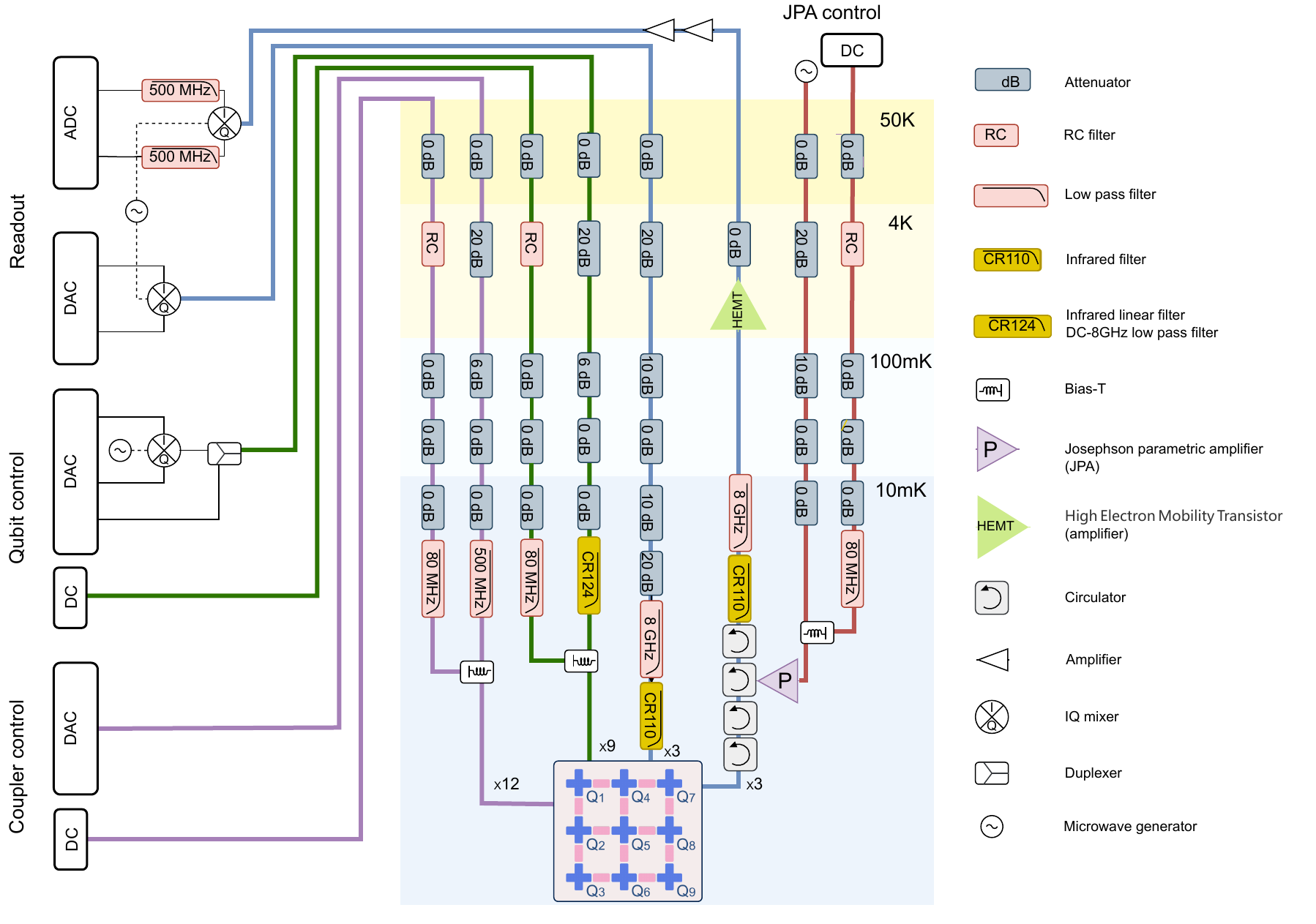}
	\caption{\small{\textbf{Experimental setup.} Electronics and wirings for
			synthesizing and transmitting the control/readout signals. Each qubit has three
			control channels: XY (microwave pulse), fast Z (Z pulse), and slow Z (direct
			current). Each coupler has two control channels: fast Z and slow Z. Readout
			pulses are generated in a similar way to the XY signals and are passed through
			the processor via the readout transmission lines. All the control and readout
			lines are well attenuated and filtered for noise shielding and precise control.
			The superconducting processor shown at the bottom has 9 qubits (crosses) and 12
			couplers (rectangles) arranged in a 3$\times$3 lattice configuration. The qubit index 
			is indicated and the coupler between Q$_i$ and Q$_j$ will be denoted as C$_{ij}$.}}
	\label{fig:wiring}
    \end{figure*}

	\medskip
	\noindent\textbf{Device information}.
	The superconducting processor used in this work contains 9 transmon qubits and 12 couplers fabricated using the flip-chip technology with a tantalum base layer~\cite{fab,fab2}. The qubits and couplers are arranged in a 3$\times$3 square lattice, as is shown schematically at the bottom in Fig.~\ref{fig:wiring}. The qubits have an anharmonicity around $-$200 MHz, and the couplers, which are also transmon qubits, have an anharmonicity around $-$130 MHz. Each qubit has a single control line for XY and Z signals, while each coupler has a flux line to adjust the effective coupling strength between the neighboring qubits. 
	
	\begin{table}[t]
		\caption{\label{tab:table1}
			\small{\textbf{Basic parameters of six qubits Q$_1$-Q$_6$ used in the present
		     experiment.} $\omega_{r}$, $\omega_{\rm{max}}$, and $\omega_{\rm{idle}}$ are
		     the readout resonator frequency, the qubit maximum frequency, and the qubit
		     idle frequency, respectively. $\alpha$ is the qubit anharmonicity. $T_{1}$
		     and $T_2^{\ast}$ are the energy relaxation time and dephasing time of the
		     qubit at idle point. $F_{g}$ and $F_{e}$  are the readout fidelities for the
		     ground and first-excited states. The infidelities of the single-qubit gate and two-qubit 
		     CZ gate (between Q$_1$ and Q$_2$, Q$_2$ and Q$_3$, Q$_4$ and Q$_5$, Q$_5$ and 
		     Q$_6$, Q$_1$ and Q$_4$, Q$_2$ and Q$_5$, Q$_3$ and Q$_6$) are calibrated by
		     RB and/or XEB. The data in parenthesis for the two-qubit CZ gate are values
		     intentionally increased for the study described in the main text. In the single-qubit case, 
		     the measurements are performed simultaneously for all qubits.}}
                \begin{ruledtabular}
                    \begin{tabular}{cccccccc}
                        Qubit& $\rm Q_1$ & $\rm Q_2$ & $\rm Q_3$&$\rm Q_4$ & $\rm Q_5$ & $\rm Q_6$   \\
                        \hline
                        $\omega_{r}/2\pi(\mathrm{GHz})$& 7.3146 & 7.3000 &7.2433 & 7.3530 & 7.3438& 7.2810 \\
                        $\omega_{\rm{max}}/2\pi(\mathrm{GHz})$& 5.142 & 5.061 & 5.051& 5.097 &4.990 & 4.973 \\
                        $\omega_{\rm{idle}}/2\pi(\mathrm{GHz})$& 5.142 & 4.907 & 5.046 & 4.967 & 4.653 & 4.792 \\
                        $\alpha/2\pi(\mathrm{MHz})$& -198 & -196 & -201 & -196 & -224 & -193  \\
                        $T_{1}(\mu {\rm s})$ & 20.4 & 46.4 & 37.8 & 25.1 & 52.28 & 25.9 \\ 
                        $T_{2}^{\ast}(\mu {\rm s})$ & 5.40 & 2.03 & 22.9 & 2.73 &1.60 & 2.03  \\
                        $F_g(\%)$ & 96.6 &94.7 & 95.7 &94.4 & 94.1&93.1 \\
                        $F_e(\%)$ & 93.8&91.4 & 94.0 & 90.2 & 90.1&92.0\\ 
                        \hline
                        1Q RB $(\%)$ & 0.049&0.038& 0.045&0.048 & 0.120&0.100\\
                        1Q XEB $(\%)$ & 0.0593&0.048 &0.043 &0.0735&0.2021 & 0.1075 \\
                        \hline
                        \hline
                        Qubits & $\rm Q_1$-$\rm Q_2$ &$\rm Q_2$-$\rm Q_3$& $\rm Q_4$-$\rm Q_5$& $\rm Q_5$-$\rm Q_6$& $\rm Q_1$-$\rm Q_4$& $\rm Q_2$-$\rm Q_5$& $\rm Q_3$-$\rm Q_6$\\
                        2Q RB$(\%)$& 0.51&0.61&(3.99)&(0.92)&0.67&(1.1)&0.80 (1.47)
                    \end{tabular}
                \end{ruledtabular}
            \end{table}

	Six qubits from Q$_1$ to Q$_6$ are used in the present experiment (see Fig.~\ref{fig:wiring}) and basic device parameters are listed in Tab.~\ref{tab:table1}.  We note that the frequency of the readout resonator $\omega_r$, ranging from 7.300 to 7.353 GHz, is above the maximum qubit frequency $\omega_{\rm{max}}$ that varies between 4.973 and 5.142 GHz. The qubit idle frequency $\omega_{\rm{idle}}$ is tuned to be close to $\omega_{\rm{max}}$ in order to reduce the effect of flux noise. The energy relaxation time $T_1$ for each qubit at the idle point is above 20 $\mu$s, and the dephasing time $T_2^{\ast}$ ranges from 1.60 to 22.9 $\mu$s. The single-qubit gate fidelities are about 99.9$\%$. The coupling strength between the qubit and the coupler connected to it is $\sim$ 80 MHz, and the direct coupling strength between two neighboring qubits is $\sim$ 9 MHz. The effective coupling strength $g$ of the neighboring qubits ranges from 1.5 to $-$50 MHz as determined by the vacuum Rabi oscillation measurement at the resonance point. The two-qubit CZ gate fidelities are $\sim$ 99.3$\%$. To achieve high performance of the single- and two-qubit gate operations, a series of signal calibrations and corrections have been done, which are detailed below.
	
	\medskip
	\noindent\textbf{Measurement setup}.
	Our experiments are performed on a BlueFors XLD-1000 dilution refrigerator with a base temperature of 10 mK. The device is placed in a superconducting sample box which is magnetically shielded in a double-layer Cryoperm-10 cylinder. The details of the measurement setup are shown in Fig.~\ref{fig:wiring}. On the left side, the purple, green, and blue lines indicate the coupler control, the qubit control, and the readout control, respectively. On the right side, the red lines represent the Josephson parametric amplifier (JPA) control. All attenuators at each cryogenic temperature stage are utilized to remove the excess thermal photons from the higher temperature stage. The XY control line and fast Z line are combined by a duplexer at room temperature, further infrared filtered by CR124, and finally combined with the DC line by a bias-Tee into the device. The combination is considered due to the fact that each qubit has only one control line in the device. Similarly, the coupler control contains a flux line and DC line, which are also combined by the bias-Tee at the lowest temperature stage. The readout control includes the input line and the output line. When the multi-tone microwave signals are sent to the transmission line via the input line, the output signals are amplified successively by the Josephson parametric amplifier, high electron mobility transistor (HEMT), and room-temperature microwave amplifiers in the output line, which are ultimately demodulated by the analog-digital converter.
	
	\medskip
	\noindent\textbf{Calibration of Z pulse distortion}. 
	Due to the impedance mismatch in the Z control line, there exist distortions with the Z pulse, such as overcharge, ringing, as well as prolonged rising and falling edges~\cite{signal}. The resulting deviation of the qubit frequency would lead to the accumulation of additional kinetic phases. Moreover, the Z pulse distortion makes the timing of the required interactions difficult to control, which is a serious problem for the realization of high-fidelity two-qubit gates. However, for a linear time-invariant system, these waveform distortions can be corrected by the method of deconvolution~\cite{ButscherJ2018}. In this way, the Z pulse distortion can be calibrated to achieve an accurate required pulse shape for the two-qubit gate implementation.
	
\begin{figure*}[t] \centering
	\includegraphics[width=0.9\textwidth]{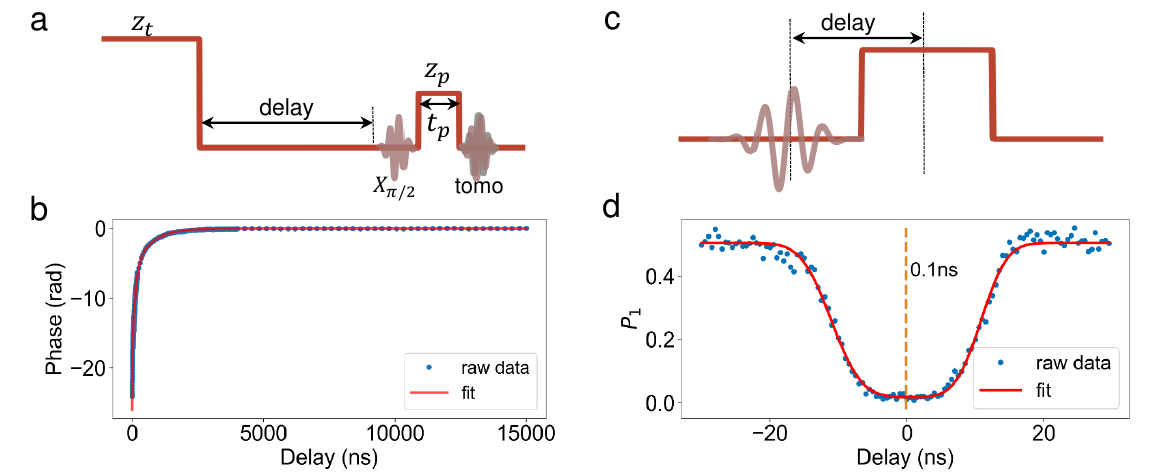}
	\caption{\small{\textbf{Calibrations of Z pulse distortion and timing of
				the single-qubit XY and Z control lines}. Subplots (a) and (b) are pulse sequence and experimental result for the calibration of Z pulse distortion, respectively. Subplots (c) and (d) are pulse sequence and experimental result for the single-qubit control line timing calibration, respectively. 
			The measurements are performed on Q$_1$.}}
	\label{fig:zdistortion}
    \end{figure*}

   The measurement and calibration of Z pulse distortion are insensitive to decoherence~\cite{Bao2022FluxoniumAA}. Figure \ref{fig:zdistortion}(a) illustrates the corresponding pulse sequence for the calibration of the qubit. The sweet point is chosen as the idle point, and the phase accumulated by the short and small Z pulse $z_p$ is used as the probe to measure the distortion of the falling edge of the long and big Z pulse $z_t$. Here $z_p$ corresponds to the flux-sensitive point. A Ramsey-like pulse sequence is used in the probing process with $z_p$ sandwiched between an $\X/2$ gate and a tomography gate. We can fit the experimental data with the above function, 
    \begin{equation}
		\delta \varphi_{t_d}
		= z_t D(z_p)\sum_{i}\tau_i a_i(e^{-(t_d+t_p)/\tau_i}-e^{-t_d/\tau_i}),
	\end{equation}
    in which $t_d$ is the delay between the falling edge and the start of the probe pulse, $\varphi_{t_d}$ is the phase accumulated by $z_p$ and $D(z_p) =\frac{\mathrm{d}\omega}{\mathrm{d}z}|_{z_p}$. Besides, $a_i$ and $\tau_i$ represent the amplitude and decay time of the exponential decay, which are obtained by fitting the experimental results. The experimental results and fits are shown in Fig.~\ref{fig:zdistortion}(b), taking Q$_1$ as an example. The fitted parameters are then used to adjust the input pulse shape such that the qubit receives an accurate required Z pulse signal. 
 
    Since the coupler does not have a readout resonator, its Z pulse distortion calibration is different from the qubit, and it needs to be read out with the help of qubit. Here we use the method in Ref.~\cite{zhao2024quantum} to achieve the Z pulse distortion calibration of the coupler by mapping the distortion of the coupler to the qubits with the help of the ac-Stark effect between the coupler and the qubit.
  	
	\medskip
	\noindent\textbf{Calibration of timing}.
	Due to the different response times of individual channels of the Z pulse and microwave pulse sources, as well as different lengths of various control lines, the time it takes for a signal to reach a qubit from a source varies among different control lines, which reduces the gate fidelity and needs to be calibrated and adjusted.
	
	\begin{figure*}[b]
		\centering
		\includegraphics[width=0.9\textwidth]{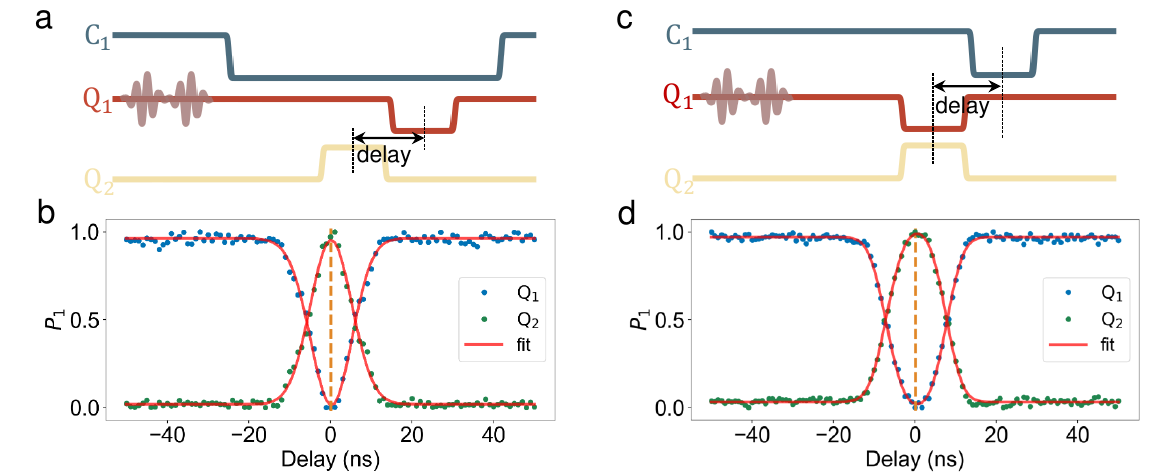}
		\caption{\small{\textbf{Timing calibrations between two neighboring qubits (a)-(b) and between a qubit and the neighboring coupler (c)-(d).} Subplots (a) and (c) are pulse sequences for the calibrations. Subplots (b) and (d) are experimental results and fits. The measurements are performed on Q$_1$, Q$_2$, and the coupler C$_{12}$ in between.}}
		\label{fig:timing}
	\end{figure*}
	
	To calibrate the timing of the XY and Z control lines of a qubit, we apply an X gate and a Z gate, and adjust the relative position of the two gates to obtain the relative timing according to the excited state probability of the qubit. Figure~\ref{fig:zdistortion}(c) shows the pulse sequence for the experiment and the result measured on Q$_1$ is shown in (d), in which a time difference of 0.1 ns between the XY and Z lines is indicated.
	
\begin{figure*}[t]
		\centering
		\includegraphics[width=0.9\textwidth]{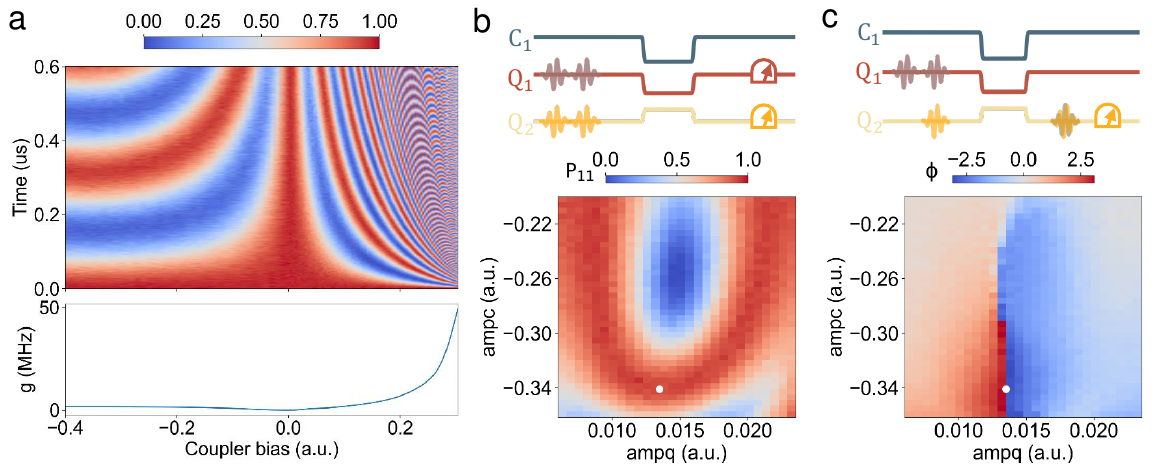}
		\caption{\small{\textbf{Implementation of the CZ gate between Q$_1$ and Q$_2$.} (a) Measurement of the effective coupling strength between Q$_1$ and Q$_2$ as a function of the coupler bias. Measurements of the leakage (b) and conditional phase (c) of the CZ gate to find the optimal points of the pulse amplitudes for qubit (ampq) and coupler (ampc). In (b) and (c), the upper parts show the pulse sequences for the measurements.}}
		\label{fig:CZ}
	\end{figure*}

	We perform the vacuum Rabi experiment to calibrate the timing between two adjacent qubits, and between a qubit and its adjacent coupler. For calibrating the timing between two qubits interleaved by a coupler, we consider the Z pulses of qubits Q$_1$, Q$_2$, and the coupler C$_{12}$ between them. The pulse sequence for calibrating the timing of the Z lines of qubits Q$_1$ and Q$_2$ is shown in Fig.~\ref{fig:timing}(a) and the experimental results are shown in (b). We prepare the initial state $|100\rangle$ (or $|001\rangle$), the Z pulse lengths of the qubits and the coupler are fixed at $t_{\rm iSWAP} =\pi/g$ and 2 $-$ 6 times of $t_{\rm iSWAP}$, respectively, here $g$ is the effective coupling strength between the two qubits (see Fig.~\ref{fig:CZ}(a) where $g$ can be tuned from 1.5 to $-$50 MHz). The Z pulses of the two qubits are moved within the time range of the coupler Z pulse to complete the timing calibration. The pulse sequence for calibrating the timing of the Z lines of qubit Q$_1$ and coupler C$_{12}$ is shown in Fig.~\ref{fig:timing}(c) and the experimental results are presented in (d). In this case, the Z pulses of the two qubits are aligned and their lengths are the same as that of the coupler. We complete the timing calibration by moving the relative timing between the coupler Z pulse and the qubit Z pulse.
	
	In the actual calibration process, the timing calibration is performed using the XY line of Q$_1$ as the reference. Timing calibration is done sequentially between the XY and Z lines of the qubit, between the Z lines of qubits Q$_1$ and Q$_2$, and between the Z lines of qubit Q$_1$ and its adjacent coupler C$_{12}$. Likewise, the calibrations can be applied to other couplers, such as Q$_2$ and Q$_3$, Q$_4$ and Q$_5$, and so on.
	
	\medskip
	\noindent\textbf{Calibration of single-qubit gates and readout}.
	The single-qubit gates including X/2 and Y/2 are realized by 20 ns microwave pulses with cosine envelopes, where the quadrature correction terms with DRAG coefficients are implemented to minimize state leakage to higher levels~\cite{DRAG}.
	We characterize the performance of the single-qubit gates by randomized benchmarking (RB) and cross-entropy benchmarking (XEB) at idle points. The measured averaged fidelities are listed in Tab.~\ref{tab:table1}.  To implement single-qubit gates with an arbitrary angle, we use $U(\theta,\phi,\lambda)=\RZ(\phi-\pi/2)\RX(\pi/2)\RZ(\pi-\theta)\RX(\pi/2)\RZ(\lambda-\pi/2)$ where $\lambda$, $\phi$ and $\theta$ are arbitrary angles and $\RZ$ is a virtual gate~\cite{PhysRevA.96.022330}. The parameterization of single-qubit gates in this way does not increase gate count and gate operation time, which is of great importance to raise the circuit fidelity to unity after RL compilation, as discussed in the main text.
	
	The duration of readout pulses for all qubits is around 1.0 $\mu$s. To achieve high-visibility and low-error readouts, the power and frequency of the readout pulses and the demodulation time are optimized. The readout fidelities of the ground and first-excited states are around 95 $\%$ and 92 $\%$, respectively, which are also presented in Tab.~\ref{tab:table1}.
	
	\medskip
	\noindent\textbf{Calibration of two-qubit CZ gate}. 
	The CZ gate between two neighboring qubits is implemented by adjusting the energy levels with an error-function shaped pulse from idle frequency to working frequency, so that the $|101\rangle$ and $|200\rangle$ (or $|002\rangle$) energy levels can be tuned to resonate for a specific amount of time.
	At the same time, the coupling is turned on to accumulate enough phase through resonance.
	
	The calibration of a CZ gate is carried out in two steps. The first step is to find the working points of the qubits and the coupler. For this we fix the CZ gate time, and measure the leakage and the conditional phase angle, as is shown in Figs.~\ref{fig:CZ}(b) and (c). To measure the leakage from the $|101\rangle$ state, we prepare the state by applying an X gate to both Q$_1$ and Q$_2$ and measure the population of the $|101\rangle$ state after applying the CZ gate. To get the conditional phase angle, we measure cross Ramsey for Q$_2$ (Q$_1$) when Q$_1$ (Q$_2$) is applied with an I or X gate, which can be used to measure the conditional accumulation of Q$_2$ (Q$_1$). We then find the optimal working points by minimizing the probability of $1-P_{|101\rangle}$ and making the conditional phase angle closest to $\phi=180^{\circ}$, as is shown in  Figs.~\ref{fig:CZ}(b) and (c) by the white dots. The Q$_1$ and Q$_2$ Z-pulse amplitudes as well as the coupler Z-pulse amplitude are then determined. The second step is to measure the single-qubit phase accumulation on Q$_1$ and Q$_2$ that accompanies the CZ gate by QPT. We subsequently apply virtual $Z$ gates to offset the phase.
	
	\begin{figure*}[t]
		\centering
		\includegraphics[width=1.0\textwidth]{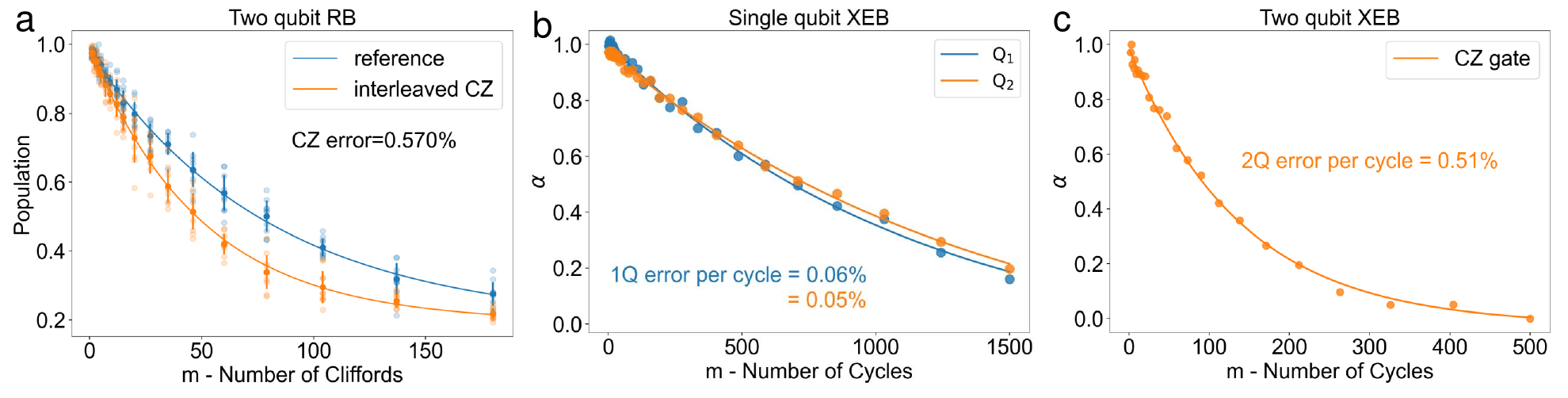}
		\caption{\small{\textbf{Comparison of RB and XEB in benchmarking the two-qubit CZ gate.} (a) Interleaved RB data to characterize the CZ gate on Q$_1$
				and Q$_2$. (b) XEB data is taken simultaneously on Q$_1$ and Q$_2$ to characterize the single-qubit gates ($\alpha$ is the sequence fidelity). (c) XEB data to characterize the CZ gate between Q$_1$ and Q$_2$, where each cycle contains two single-qubit gates in parallel and a CZ gate. The CZ Pauli errors extracted from RB and XEB are
				0.57$\%$ and 0.51$\%$, respectively.}}
		\label{fig:rbxeb}
	\end{figure*}
	
	Based on the above method, we are able to implement the CZ gates between seven qubit pairs, whose parameters are listed in Tab.~\ref{tab:table1}. The total CZ gate time is 55 ns, where both the rising and falling edges of the pulse are 6 ns, the flattening part for the accumulating phase is 39 ns (corresponding to $g/2\pi \approx$  $-$9 MHz), and the buffer time is 4 ns. In Fig.~\ref{fig:rbxeb}, the characterizations of the CZ gate between $\rm Q_1$ and $\rm Q_2$ by randomized benchmarking (RB) and cross-entropy benchmarking (XEB)~\cite{Quantumsupremacy} at idle points are characterized, with consistent Pauli errors. As an example, the fidelities of the CZ gates between Q$_1$ and Q$_2$ and between Q$_1$ and Q$_4$,  characterized by XEB, are 99.49 $\%$ and 99.33 $\%$, respectively.
	
	\medskip
	\noindent\textbf{Cross entropy benchmarking (XEB)}.
    We characterize the single-qubit and two-qubit gates (Fig.~\ref{fig:rbxeb}), as well as the performance of compiled circuits by XEB. For single-qubit gates, each cycle in an XEB circuit consists of a $\pi/2$ rotation randomly chosen from the following set: $\{\R_{\phi}(\pi/2)\big|\phi\in \{0,\pi/4,\pi/2,3\pi/4,\pi,5\pi/4,3\pi/2,7\pi/4\}\}$, where $\phi$ represents a specific angle. The rotation operator itself can be expressed as $\text{R}_{\phi}(\pi/2)=\exp({-i\frac{\pi}{4}(\cos{\phi}\text{X}+\sin{\phi}\text{Y})})$. For two-qubit sequences, each cycle contains a layer of two single-qubit gates $\R_{\phi}(\frac{\pi}{2})$ followed by a CZ gate. We refer to Ref.~\cite{CharacterizingQuantumSupremacy} for the calculation of fidelity using measured probabilities.

	\medskip
    \noindent\textbf{Quantum process tomography (QPT)}.
    We characterize the performance of compiled circuits running on the quantum processor by QPT~\cite{PhysRevA.64.012314}. 
    We select $\bigotimes^n\{\text{I}, \X, \Y/2, \X/2 \}$ to prepare the initial state, and select $\otimes^n \{-\Y/2, \X/2, \text{I}\}$ to complete the projection measurement of quantum states in different axial directions. In order to suppress fluctuations caused by experimental errors, each measurement is repeated 4000 times. The experimental fidelity is obtained by $F_2(\hat{U}, \hat{V}) = \Tr ( \mathcal{\hat{E}}^{\dagger} \hat{\mathcal{F}})$, where $\mathcal{\hat{E}}$ and $\hat{\mathcal{F}}$ are the calculated and measured process matrices of the target unitary $\hat{U}$ and the unitary $\hat{V}$ of the compiled circuit, respectively. 

\vspace{10mm}
\section{RL-BASED QUANTUM COMPILER}\label{appendix:scheme_detail}
\vspace{5mm}

	\medskip
	\noindent\textbf{Quantum compiling.} For quantum compiling with discrete native basis set $\mathcal{G}$ and given the topology of the quantum processor, we define the generalized basis set $\tilde{\mathcal{G}}$ whose element corresponds to the executable operation on the processor. For example, for a three-qubit system with chain structure Q$_1$--Q$_2$--Q$_3$, if $\X$ gate is in $\mathcal{G}$, then the generalized basis gate $\tilde{\mathcal{G}}$ contains $\mathbb{I}_4\otimes \X$, $\X\otimes \mathbb{I}_4$, and $\mathbb{I}_2\otimes \X \otimes \mathbb{I}_2$. Similarly, when $\CZ\in \mathcal{G}$, the corresponding elements in $\tilde{\mathcal{G}}$ are $\CZ\otimes\mathbb{I}_2$ and $\mathbb{I}_2\otimes\CZ$. This generalized basis set $\tilde{\mathcal{G}}$ is dubbed action space in the context of RL-based quantum compiler, as denoted by  $\mathcal{A}=\{A_i\}_{i=1}^d$. 
	
	The purpose of quantum compiling is to find a sequence of quantum gates $\{A_1', A_2', \dots, A_L'\}\in \mathcal{A}^N$ with the minimum length $N$ such that the distance  between the target unitary $\hat{U}$ and $\hat{V}=\prod_{j=1}^{N} A_{j}'$ is bounded by a pre-defined error $\varepsilon$, i.e., 
	\begin{eqnarray}\label{eqn:QCL-def}
		d\left(\hat{V},\hat{U}\right)<\varepsilon,
	\end{eqnarray}
	where the distance metric $d(\cdot, \cdot)$  refers to two infidelity measures $1-F_1(\hat{V},\hat{U})$ and $1-F_2(\hat{V},\hat{U})$ defined in the main text. 
	
	\medskip
	\noindent\textbf{Training data Generation}. 
	Here we elucidate the omitted construction rule of training data in Methods. Recall that the cardinality of $\mathcal{W}_l$ exponentially scales with $l$, which may lead to memorization and computation issues for a large $l$. To tackle this issue, a threshold value $l^*$ is introduced. When $l\leq l^*$, the percepts $\mathcal{W}_l$ are generated in a brute force manner; when $l>l^*$, the percepts are generated in a random sampling manner. The setting of the threshold $l^*$ is heuristic, depending on the available memory resources, the computation power, and the maximum loop number $L$. We refer to Supplementary Information~\ref{appendix:two-qubit-res} and \ref{appendix:three-qubit-res} for the detailed setting of $l^*$. 

	\medskip
	\noindent\textbf{RL-based quantum compiler}. 
	The pseudo-code of DQN employed in our proposal is presented in Alg.~\ref{alg:train_DQN}. To improve the training stability, we adopt the technique developed in  Refs.~\cite{mnih2015human, agostinelli2019solving} to optimize DQN. Concretely, following notations in the main text, the loss function at the $l$-th loop to be minimized takes the form
	\begin{eqnarray}\label{eqn:DQN_loss}
		\mathcal{L}_{l}(\bm{\theta}) = \frac{1}{|\mathcal{W}_l|\cdot |\mathcal{A}|} \sum_{W_i\in\mathcal{W}_l} \sum_{A\in\mathcal{A}} \left( Q_{\bm{\theta}}(W_i, A)-\left( \max_{A^{\prime}\in\mathcal{A}}Q_{\bm{\theta}^{-}}(AW_i, A^{\prime}) -1 \right) \right)^{2} ,
	\end{eqnarray}
	where $W_i\in\mathcal{W}_l$, $Q_{\bm{\theta}}$ represents a DQN  and $Q_{\bm{\theta}}(W_i, A)$ represents its estimation value of the percept $AW_i$, and $Q_{\bm{\theta}^{-}}$ represents an auxiliary DQN as elaborated below. 
	
	Compared to the loss defined in Methods, an auxiliary DQN $Q_{\bm{\theta}^{-}}$ is introduced in Eq.~(\ref{eqn:DQN_loss}) to replace $\hat{V}_{ji}$. The two DQNs $Q_{\bm{\theta}^{-}}$ and $Q_{\bm{\theta}}$ have identical structures and similar parameters. The only difference is that the original DQN $Q_{\bm{\theta}}$  dynamically updates $\bm{\theta}$ using the back-propagation, while the auxiliary DQN $Q_{\bm{\theta}^{-}}$ stays intact most of the time and provides the estimation of the target values, i.e., $\max_{A^{\prime}\in\mathcal{A}}Q_{\bm{\theta}^{-}}(AW, A^{\prime}) -1$.
	
	The optimization of $\mathcal{L}_l(\bm{\theta})$ in Eq.~(\ref{eqn:DQN_loss}) can be completed by the gradient descent methods. For each $l\in[L]$, the training parameters $\bm{\theta}$ of the original DQN are updated up to $T$ epochs (Line 3 in Alg.~\ref{alg:train_DQN}). An early termination is allowable if the loss $\mathcal{L}(\bm{\theta})$ is below a threshold $\delta$ (Line 8 in Alg.~\ref{alg:train_DQN}). Then, we move to the next loop $l\rightarrow l+1$ and the parameters $\bm{\theta}^-$ are updated to $\bm{\theta}$.    
	
	\begin{algorithm}[h!]
		\caption{Training DQN assisted by the auxiliary DQN}
		\label{alg:train_DQN}
		\DontPrintSemicolon
		\KwIn{Action space $\mathcal{A}$; Loss threshold $\delta$; Adam optimizer with the initial learning rate $\gamma$; The maximum loop $L$; Epoch $T$; Initialized the original and target DQNs with the parameters ${\bm{\theta}}$ and ${\bm{\theta}^-}$; Threshold $\delta$.}
		\KwOut{The trained target DQN.}
		\BlankLine
		Set $l=1$\;
		\While{$l \leqslant L$}{
			Generate the training data $\mathcal{W}_{l}$ \;
			\For{$i\leftarrow 1$ \KwTo $T$}{
				Sample random minibatch of percepts from $\mathcal{W}_{l}$ and $\mathcal{A}$ \;
				Calculate the loss $\mathcal{L}_{l}(\bm{\theta})$ in Eq.~(\ref{eqn:DQN_loss}) \;
				Compute the gradient $\nabla_{\bm{\theta}}\mathcal{L}_{l}(\bm{\theta})$ \;
				Update $\bm{\theta}$ based on the gradients \;
				\If{loss $\leqslant \delta$}{
					$\bm{\theta}^{-}\leftarrow \bm{\theta} $ \;
					$l \leftarrow  l+1$
				} 
			}
			$\bm{\theta}^{-}\leftarrow \bm{\theta} $ \;
			$l \leftarrow  l+1$
		}
	\end{algorithm}
	
	\medskip
	\noindent\textbf{Multi-DQN collaborative AQ* search}. We have seen in the main text that DQNs cannot learn the whole unitary space when the total number of loops $L$ is limited. To improve the RL-based compiler in the inference stage, we adapt the original AQ* search (Alg.~\ref{alg:AQSearch}) to the multi-DQN collaborative version. The pseudo-code of the multi-DQN collaborative AQ* search algorithm is summarized in Alg.~\ref{alg:MAQSearch}. 
	
	To better understand Alg.~\ref{alg:MAQSearch}, we consider $J$ DQN models ($J=3$ in the presented example). These models are denoted as $Q_{\bm{\theta}_1}$, $Q_{\bm{\theta}_2}$ and $Q_{\bm{\theta}_3}$, with each model being trained on distinct generalized basis gate sets: $\mathcal{A}_1$, $\mathcal{A}_2$ and $\mathcal{A}_3$, respectively. The algorithm leveraging these DQN models for collaborative AQ* search aims to find a gate sequence to approximate a target unitary matrix $\hat{U}$ while adhering to an accuracy threshold of $\varepsilon$. The core idea is a sequential collaboration between the DQN models. Initially, we use $Q_{\bm{\theta}_1}$ to execute the standard AQ* search algorithm (Alg.~\ref{alg:AQSearch}) and generates an initial guess, $\hat{V}_1$, to approximate $\hat{U}$. Subsequently, $Q_{\bm{\theta}_2}$ takes over, using the AQ* search to compile the product $\hat{V}_1^{\dagger}\hat{U}$ and derive $\hat{V}_2$. This process iterates, with each DQN model $Q_{\bm{\theta}_j}$ ($j=1, 2, 3$ in this case) focusing on a specific stage of compiling the overall sequence. The final collaborative compiling target is defined as $\hat{V}=\hat{V}_1\cdot\hat{V}_2\cdot\hat{V}_3$, where each component originates from a different basis gate set. Intuitively, this collaborative approach can enhance the accuracy of the compiled sequence at the expense of increased resources.
 
	\normalem
	\begin{algorithm}
	   \caption{AQ* Search}
	   \label{alg:AQSearch}
	   \DontPrintSemicolon
	   \KwIn{
		   Target unitary matrix $\hat{U}$; The DQN model $Q_{\bm{\theta}}$; The error tolerance $\varepsilon$; The search depth threshold $D$; The action space $\mathcal{A}$; The initial percept $\mathbb{I}$.
		   }
	   \KwOut{Compiling sequence $\hat{V}$ to approximate $\hat{U}$.}
	   \BlankLine

          Set $\hat{V}=\mathbb{I}$ \;
          \For{\underline{$i\leftarrow 1$ \KwTo $D$}}{
          $A^{\star}=\arg \max_{A \in \mathcal{A}} f(\hat{V}, A ; \hat{U})$ \;
          Execute the matrix multiplication $\hat{V} = \hat{V}\cdot A^{\star}$ \;
          \If{\underline{$F_1(\hat{V}, \hat{U})\leq \varepsilon$}}{
          \Return $\hat{V}$
          }
          }
          \Return $\hat{V}$
	\end{algorithm}
	
	\normalem
	\begin{algorithm}
		\caption{Multi-DQN collaborative AQ* search}
		\label{alg:MAQSearch}
		\DontPrintSemicolon
		\KwIn{Target unitary matrix $\hat{U}$; DQN models $Q_{\bm{\theta}_1}$, $Q_{\bm{\theta}_2}$ $\cdots$ $Q_{\bm{\theta}_J}$; The error tolerance $\varepsilon$; The search depth threshold $D$; The action space $\mathcal{A}$; The initial percept $\mathbb{I}$.
		}
		\KwOut{Compiling sequence $\hat{V}$ to approximate $\hat{U}$.
		}
		\BlankLine
            Set $\hat{V}=\mathbb{I}$ \;
          \For{\underline{$j\leftarrow 1$ \KwTo $J$}}{
          $\hat{V}$ = \text{AQ* Search}($\hat{U}$, $Q_{\bm{\theta}_j}$, $\varepsilon$, $D$, $\mathcal{A}$, $\hat{V}$) \;
          \If{\underline{$F_1(\hat{V}, \hat{U})\leq \varepsilon$}}{
          \Return $\hat{V}$
          }
          }
          \Return $\hat{V}$
	\end{algorithm}
	
	\medskip
	\noindent\textbf{Comparison with other RL-based compilers}. Two previous works ~\cite{zhang2020topological,moro2021quantum} have utilized the RL techniques for quantum compilation, showing varying performance in accomplishing the quantum compiling tasks. The proposal in Ref.~\cite{moro2021quantum} involves policy-based RL, which encounters the sparse reward problem. To address the issue, our proposal utilizes value-based RL, which can be applied to any universal basis set and multi-qubit systems. While the proposal in Ref.~\cite{zhang2020topological} also belongs to the value-based RL, it requires prior knowledge in compiling two-qubit gates to ensure performance in terms of fidelity. Our proposal removes this requirement by adopting a more powerful model, DQN, and a different loss function. As a result, it allows for a shorter gate sequence length and a shorter inference time in compiling multi-qubit circuits.
	
\vspace{10mm}
\section{EXTENDED DISCUSSION AND RESULTS FOR TWO-QUBIT SYSTEM}\label{appendix:two-qubit-res}
\vspace{5mm}

	In this section, we describe the construction rule of the training data and test data, and the implementation details of the RL-based compiler, followed by the presentation of more numerical and experimental results of compilation for the two-qubit system. Finally, a summary of the results is presented.
	
	\subsection{Construction of training data and test data}\label{appendix:subC1:data-generation}
	
	\noindent\textbf{Data generation}. We exploit three basis sets for different two-qubit compiling tasks. The first basis set is $\mathcal{G}^{(2,1)}= \{ \pm \X/2,\pm \Y/2, \T, \T^{\dagger}, \CZ \}$, where the superscript $(2,1)$ represents the two-qubit case under the first basis set. Accordingly, the generalized basis set, or equivalently the action space, yields
	\begin{equation}
		\begin{aligned}
			\tilde{\mathcal{G}}^{(2,1)} \equiv \mathcal{A}^{(2,1)}=\Big\{& \RX(\frac{\pi}{2})\otimes \mathbb{I}_2, \mathbb{I}_2 \otimes \RX(\frac{\pi}{2}), \RX(-\frac{\pi}{2})\otimes \mathbb{I}_2, \mathbb{I}_2\otimes \RX(-\frac{\pi}{2}),   \RY(\frac{\pi}{2})\otimes \mathbb{I}_2, \mathbb{I}_2\otimes \RY(\frac{\pi}{2}), \\ 
			& \RY(-\frac{\pi}{2})\otimes \mathbb{I}_2, \mathbb{I}_2\otimes \RY(-\frac{\pi}{2}),  T\otimes \mathbb{I}_2, \mathbb{I}_2\otimes T, T^{\dagger}\otimes \mathbb{I}_2, \mathbb{I}_2\otimes T^{\dagger}, \CZ \Big\},
		\end{aligned}
	\end{equation}
	with the cardinal number $|\mathcal{A}^{(2,1)}|=13$. 
	
The second basis set is $\mathcal{G}^{(2,2)}= \{ \pm \X/6,\pm \Y/6, \pm \Z/6 \}$, where $\rm P/6$ represents rotation gate $\rm R_P(\pi/6)$ for $\rm P\in\{X, Y, Z\}$. Accordingly, the action space with respect to the second basis set yields
	\begin{equation}
		\begin{aligned}
			\mathcal{A}^{(2,2)}= \Big\{& \RX(\frac{\pi}{6})\otimes \mathbb{I}_2, \mathbb{I}_2 \otimes \RX(\frac{\pi}{6}), \RX(-\frac{\pi}{6})\otimes \mathbb{I}_2, \mathbb{I}_2\otimes \RX(-\frac{\pi}{6}),  \RY(\frac{\pi}{6})\otimes \mathbb{I}_2, \mathbb{I}_2\otimes \RY(\frac{\pi}{6}), \RY(-\frac{\pi}{6})\otimes \mathbb{I}_2, \mathbb{I}_2\otimes \RY(-\frac{\pi}{6}), \\ 
			& \RZ(\frac{\pi}{6})\otimes \mathbb{I}_2, \mathbb{I}_2\otimes \RZ(\frac{\pi}{6}), \RZ(-\frac{\pi}{6})\otimes \mathbb{I}_2, \mathbb{I}_2\otimes \RZ(-\frac{\pi}{6}) \Big\},
		\end{aligned}
	\end{equation}
	with the cardinal number $|\mathcal{A}^{(2,2)}|=12$. 
	
The third basis set is $\mathcal{G}^{(2,3)}= \{ \pm \X/128,\pm \Y/128, \pm \Z/128 \}$, where $\rm P/128$ represents the rotation gate $\rm R_P(\pi/128)$ for $\rm P\in\{X, Y, Z\}$. Accordingly, the action space with respect to the third basis set yields
	\begin{equation}
		\begin{aligned}
			\mathcal{A}^{(2,3)}=\Big\{& \RX(\frac{\pi}{128})\otimes \mathbb{I}_2, \mathbb{I}_2 \otimes \RX(\frac{\pi}{128}), \RX(-\frac{\pi}{128})\otimes \mathbb{I}_2, \mathbb{I}_2\otimes \RX(-\frac{\pi}{128}),   \RY(\frac{\pi}{128})\otimes \mathbb{I}_2, \mathbb{I}_2\otimes \RY(\frac{\pi}{128}), \\ 
			&  \RY(-\frac{\pi}{128})\otimes \mathbb{I}_2, \mathbb{I}_2\otimes \RY(-\frac{\pi}{128}), \RZ(\frac{\pi}{128})\otimes \mathbb{I}_2, \mathbb{I}_2\otimes \RZ(\frac{\pi}{128}), \RZ(-\frac{\pi}{128})\otimes \mathbb{I}_2, \mathbb{I}_2\otimes \RZ(-\frac{\pi}{128}) \Big\},
		\end{aligned}
	\end{equation}
	with the cardinal number $|\mathcal{A}^{(2,3)}|=12$.
	
	As explained in Sec.~\ref{appendix:scheme_detail} and Methods, the initial percept $\mathbb{I}$ is perturbed to enhance the algorithmic robustness. In the two-qubit compiling tasks, the perturbed percept takes the form
	\begin{equation}
		\mathbb{I}^{\prime} = \left(\cos{(\frac{\alpha}{2})}\mathbb{I}_2-i\sin{(\frac{\alpha}{2})} (j_1^{(1)}\X+j_2^{(1)}\Y+j_3^{(1)}\Z )\right)\otimes \left(\cos{(\frac{\alpha}{2})}\mathbb{I}_2-i\sin{(\frac{\alpha}{2})} (j_1^{(2)}\X+j_2^{(2)}\Y+j_3^{(2)}\Z) \right),
	\end{equation}
	where the tunable parameters $j_1^{(1)}, j_2^{(1)}, j_3^{(1)}, j_1^{(2)}, j_2^{(2)}, j_3^{(2)} \in \mathbb{R}$ satisfy $|j_1^{(1)}|^2+|j_2^{(1)}|^2+|j_3^{(1)}|^2=1$ and $|j_1^{(2)}|^2+|j_2^{(2)}|^2+|j_3^{(2)}|^2=1$, and $\alpha$ is randomly sampled from $[-0.002, 0.002]$.
	
	In generating the training data, the maximum loop number is set as $L=50$ and the threshold for the number of loops is set as $l^*=3$.
	
\begin{figure}[t]
		\centering
		\includegraphics[width=0.99\textwidth]{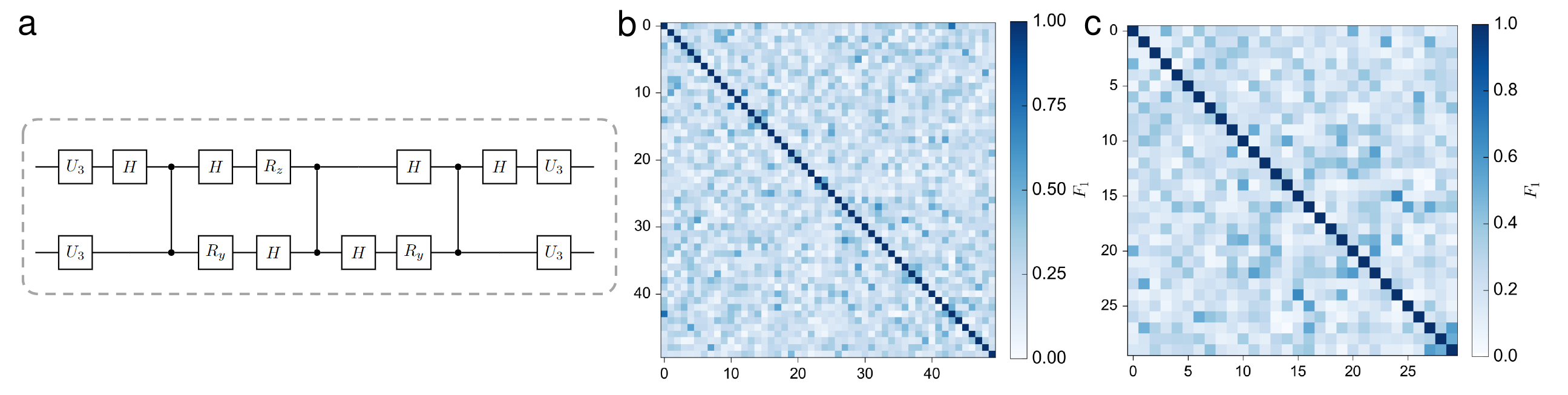}
		\caption{\small{\textbf{Generation of compiling targets.} (a) The KAK decomposition of a template circuit. (b) The heat-map for the generated $50$ target unitaries in the robustness verification task (Task 1). The correlation between compiling targets $U_1$ and $U_2$ is calculated by $F_1(U_1, U_2)$. Both the horizontal and vertical coordinates represent the target index. (c) The heat-map for the $30$ targets sampled from SU(4) unitaries in the task of compiling general unitary (Task 2).
		}}
		\label{fig:SM_sample_correlation}
	\end{figure}

	\medskip
	\noindent\textbf{Compiling targets generation}. To verify the versatility of our RL-based compiler, we benchmark its performance via the following four tasks. In Task 1, we construct $50$ target operations. Each target is constructed by randomly sampling $83$ gates from $\mathcal{G}^{(2,1)}$ and deployed on a two-qubit circuit in a specific order. Specifically, the compiling targets are generated guided by the KAK decomposition method as shown in Fig.~\ref{fig:SM_sample_correlation}(a). We collect the target unitaries by fixing three CNOT gates in a quantum circuit and operating those single-qubit gates. And $80$ single-qubit gates uniformly sampled from $\mathcal{G}^{(2,1)}$ are employed to form the single-qubit gates. Fig.~\ref{fig:SM_sample_correlation}(b) exhibits the dissimilarity of the unitaries. In Task 2, 30 random targets are sampled from SU(4). Fig.~\ref{fig:SM_sample_correlation}(c) exhibits the dissimilarity of these targets. In Task 3, we compile the $\RZZ$ gates. In Task 4, we simplify the targets in Task 1 by sampling only 43 gates from $\mathcal{G}^{(2,1)}$ to construct the targets with the same circuit structure.
	
    \subsection{Implementation of RL-based compiler}\label{appendix:subC2:implentation-details}
	
	\noindent\textbf{Classical Device}. For the proposed RL-based compiler, both the precompilation and inference processes related to the classical part are implemented on an Nvidia Tesla V100 GPU.  
	
	\begin{figure*}[htbp]
		\centering
		\includegraphics[width=0.99\textwidth]{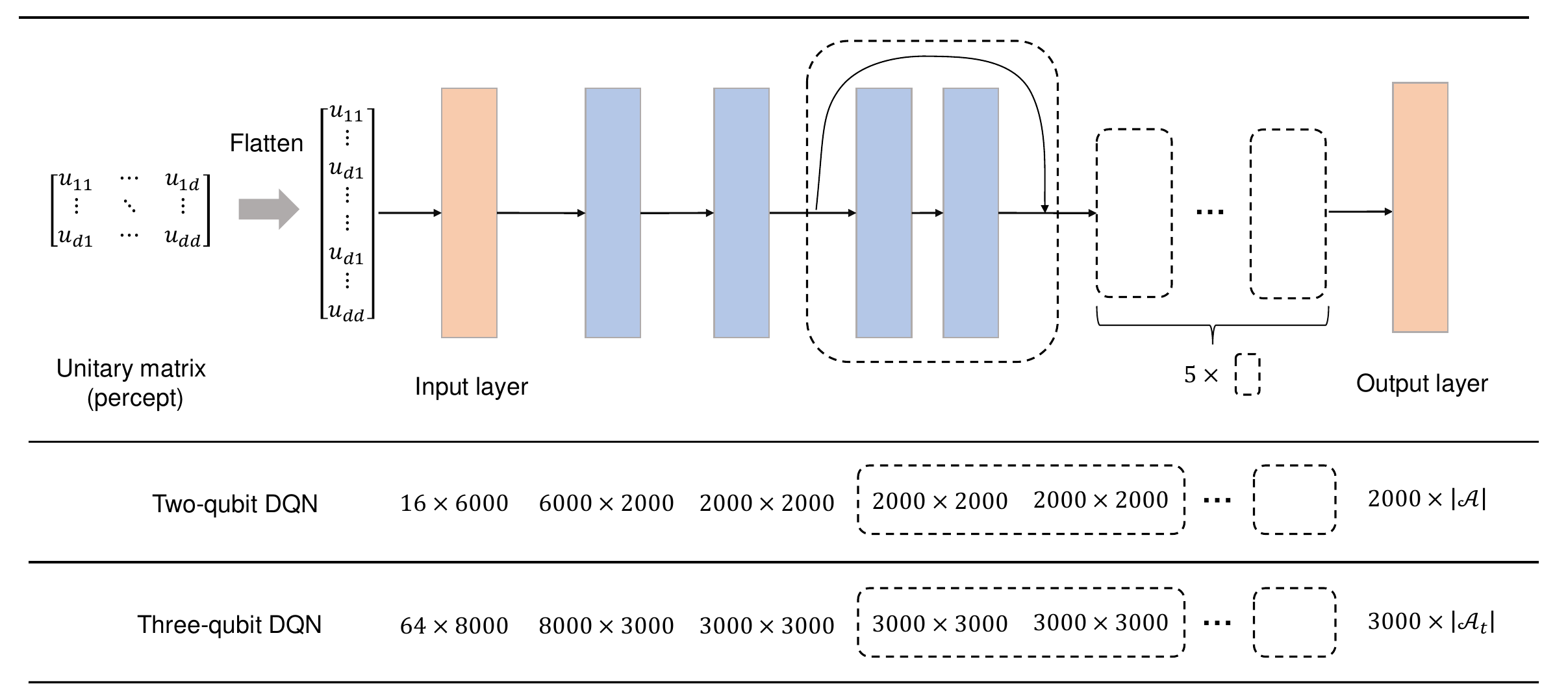}
		\caption{\small{\textbf{The structure of the employed DQN model.} The blue blocks represent the hidden layers. Every two layers are fully connected.  The output dimension of the three-qubit DQN is determined by the dimension of the action space $\mathcal{A}$.
		}}
		\label{fig:dqn_structure}
	\end{figure*}
	\medskip
	\noindent\textbf{Architecture of DQNs.} The visualization of the employed DQN is shown in Fig.~\ref{fig:dqn_structure}, which is adapted from ResNet ~\cite{he2016deep}. Specifically, it consists of an input layer, two hidden layers, and six residual blocks,  followed by $|\mathcal{A}^{(2, t)}|$ output neurons with $t\in\{1, 2, 3\}$. The input layer is of size $16$. The first two hidden layers are of sizes 6000 and 2000, and each residual block is composed of two hidden layers with 2000 hidden neurons, with the full connection. The leaky ReLU activation function and batch normalization are applied to all hidden layers.  
	
	\medskip
	\noindent\textbf{Hyper-parameters}. 
	We use Adam optimizer to optimize DQN, where the learning rate is $\eta =10^{-3}$ without the weight decay ~\cite{kingma2017adam}. The loss threshold $\delta$ in Alg.~\ref{alg:train_DQN} is set as $0.02$. The epoch $T$ depends on $l$, i.e., $T=100\cdot l$.	In the training process, the optimized DQN models at each loop $l$ under each action space $\mathcal{A}^{(2,t)}$, denoted by $\{Q^{(t, {l})}_{\bold{\theta}}\}$, are stored to conduct multi-DQN collaborative AQ* search. In the inference stage, we execute the multi-DQN collaborative AQ* search guided by DQN model $\{Q^{(t, {l})}_{\bold{\theta}}\}_{t,l}$. And the maximum AQ* search depth is set as 200 and the inference time is restricted to be less than 10 minutes.
	
	\subsection{More numerical and experimental results}\label{appendix:subC3:more-results}
	
	In this subsection, we provide more results in compiling two-qubit circuits to further evaluate the performance of the RL-based compiler in three metrics, i.e., fidelity, gate sequence length, and inference time (variational RL-based compiler will be discussed in Sec~\ref{sm_last_sebsec_vqa}). We focus on exploring the capability of the compiler and compare it with the KAK decomposition under four tasks discussed below, where the KAK decomposition is implemented by Qiskit compiler~\cite{mckay2018qiskit} whose basis gate set contains $\text{U}_\text{3}$ gate and $\CZ$ gate. As shown in Fig.~\ref{fig:SM_sample_correlation}(a), any two-qubit quantum gate can be decomposed into three $\CZ$ gates and multiple single-qubit gates. We perform experiments for the four tasks and each decomposed gate sequence is performed five times repeatedly to evaluate the statistical performance.
	
	We point out that the Qiskit compiler is mainly used as a reference for convenience to assess the performance of the RL-based compiler, and we refer to other compilers that may outperform Qiskit compiler in some metrics such as those employing heuristic search~\cite{davis2019heuristics,PhysRevX.11.041039,lu2023QuantumCompilingVariational}. Our results show that the experimental fidelity $F_2$ by RL-based compiler can possibly be higher than that by Qiskit compiler even when the corresponding circuit fidelity $F_1$ by the former (latter) is lower (higher). This is unique for the NISQ processor reflecting the interplay among circuit fidelity, circuit length and structure, as well as gate errors and decoherence of the processor, which is also seen in the discussion of the three-qubit QFT circuits in the main text.
	
	\begin{figure*}[t]
	\centering
	\includegraphics[width=1.00\textwidth]{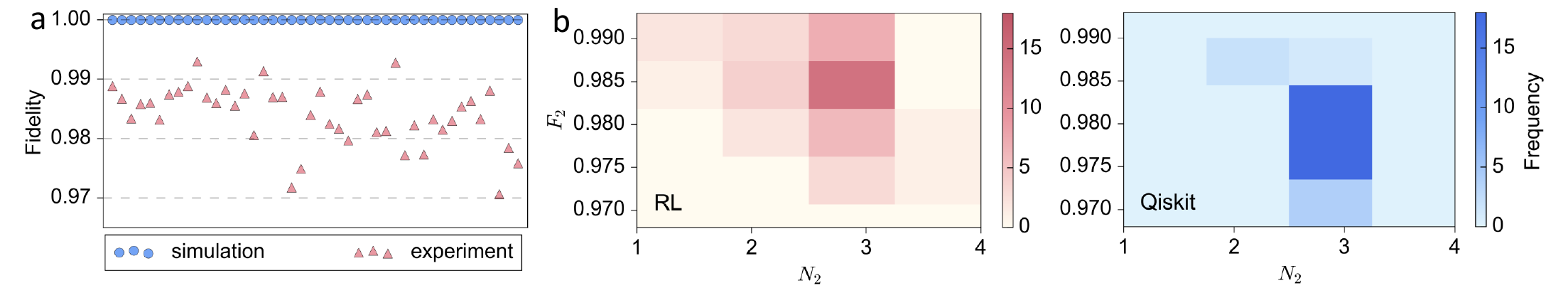}
	\caption{\small{\textbf{Compilation of random targets in two-qubit system.} (a) Simulation ($F_1$) and experimental ($F_2$) fidelity distributions of circuits by RL-based compiler for 50 target unitaries. The abscissa representing the index of targets is not shown. (b) Frequency under variations of experimental fidelity $F_2$ and two-qubit gate count $N_2$ for circuits by RL-based compiler (left panel) and Qiskit compiler (right panel). Compared with the Qiskit compiler,  the RL-based compiler requests fewer two-qubit gates and achieves a higher fidelity on average.}}
	\label{fig:SM_2_qubit_special}
    \end{figure*}

	\medskip	
	\noindent\textbf{Task 1: Compiling random target operations}. In the main text, we have shown the results of compiling the SWAP gate and $\RZZ(\pi/2)$ gate. Here we explore the performance of the RL-based compiler for more general random target operations. As can be seen in Fig.~\ref{fig:SM_2_qubit_special}, the RL-based compiler outperforms the Qiskit compiler in terms of experimental fidelity $F_2$ and two-qubit gate count $N_2$. On average, the RL-based compiler has $N_2 = 2.80$ gates to accomplish the compiling task with $F_2=0.9840 \pm 0.0049$, while the Qiskit compiler has $N_2 = 2.96$ gates with $F_2= 0.9790 \pm 0.0041$. However, the inference time of the RL-based compiler is much longer than that of the Qiskit compiler, i.e., $250$ versus $0.02$ seconds. The long inference time is caused by the constant calls to the DQN. Exploring ways to reduce the inference time is left for future work. 
	
	\medskip	
	\noindent\textbf{Task 2: Compiling random operations from SU(4)}.
	We specify the universal basis set $\mathcal{G}^{(2,2)}$ and use it to train three DQNs. The only difference between the three DQNs is the varied total loop number $L \in \{20, 25, 45\}$. The trained DQNs are applied to compile $50$ two-qubit gates whose construction rule is identical to that in Task~1. To investigate the performance of the RL-based compiler for more general two-qubit gates, we employ the trained DQNs with universal basis set $\mathcal{G}^{(2,1)}$ to compile $30$ two-qubit gates randomly sampled from SU(4). 
	
	\begin{figure}[b]
		\centering
		\includegraphics[width=0.7\textwidth]{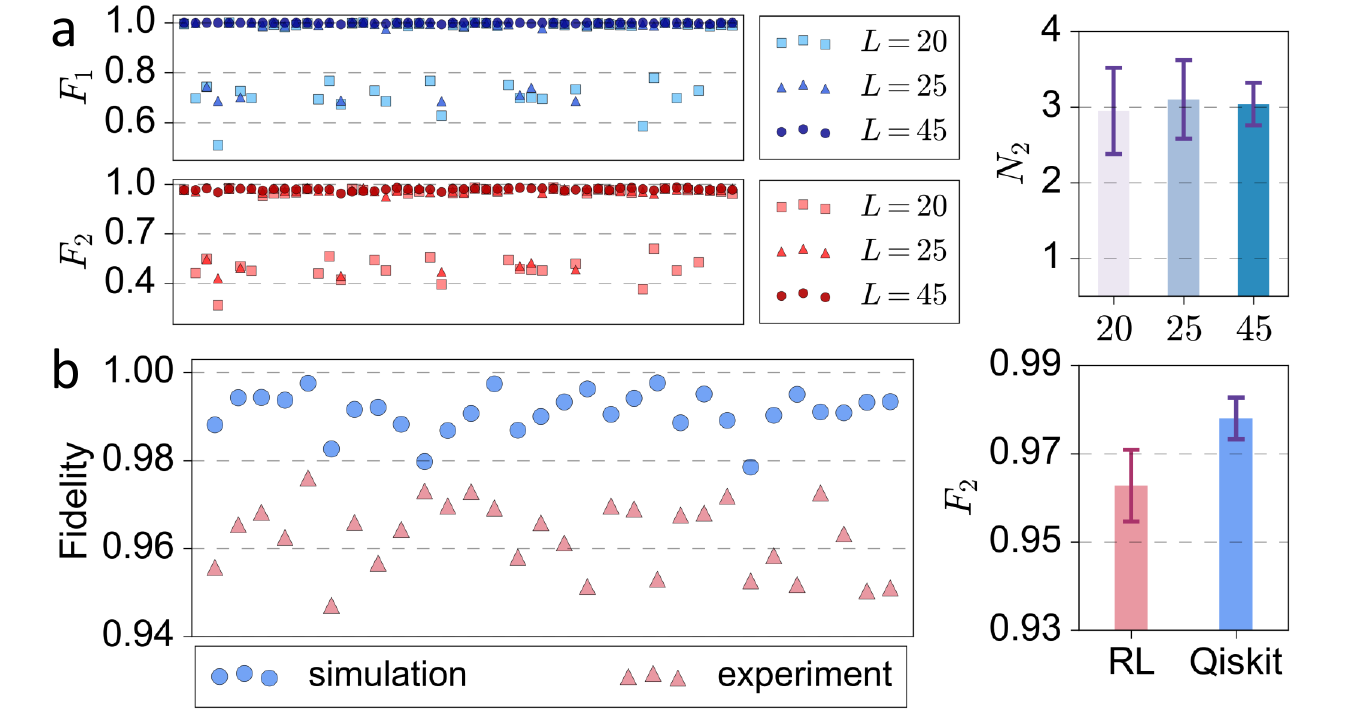}
		\caption{\small{\textbf{Scaling behavior of DQN and compilation of random operations from SU(4).} (a) Compilation of 50 two-qubit gates with restricted complexity. The left panel illustrates the performance of the RL-based compiler in the metric of $F_1$ and $F_2$ when the loop number is set as $L=20$, $25$, and $45$, respectively. The right panel shows the number of two-qubit gates $N_2$ in the output gate sequence with different $L$. The corresponding number of single-qubit gates $N_1$ on average is $10.69$, $16.96$, and $25.68$,  respectively. The inference time in all cases is around $250$ seconds. (b) Compilation of 30 two-qubit gates sampled from SU(4). The left panel shows the performance of the RL-based compiler with $L=45$ by $F_1$ (simulation) and $F_2$ (experiment). The right panel presents the experimental fidelity $F_2$ of RL-based and Qiskit compilers. The abscissa in the left panels representing the index of targets is not shown.}}
		\label{fig:exp_two_qubits_continuous_main_text}
	\end{figure}
	
	The compiling and experimental results are depicted in Fig.~\ref{fig:exp_two_qubits_continuous_main_text}. Figure~\ref{fig:exp_two_qubits_continuous_main_text}(a) indicates that the compiling accuracy of the RL-based compiler is steadily improved by increasing $L$, in the measures of both $F_1$ and $F_2$. This observation suggests that training DQN with a sufficiently large $L$ can promise a satisfactory performance. Besides, the rightmost panel shows that $N_2$ is independent with $L$, implying that the gate sequence length can be well controlled for large $L$, as a desired feature for NISQ devices. Figure~\ref{fig:exp_two_qubits_continuous_main_text}(b) illustrates the results of compiling random unitaries sampled from SU(4). On average, the RL-based compiler produces results less satisfactory compared to the Qiskit compiler in terms of $F_2$,  caused by the fact that the trained DQN with $L=45$ is insufficient to explore the whole unitary space. In practice, this gap can be mitigated by employing the variational RL-based compiler as detailed in Sec~\ref{sm_last_sebsec_vqa}.
	
	\medskip	
	\noindent\textbf{Task 3: Compiling $\RZZ(\gamma)$ gate with different $\gamma$}. In the main text, we have presented the compiling and experimental results with $\gamma=\pi/2$. Here we present the results for arbitrary angles with \[\gamma \in \left\{ \frac{5\pi}{180}, \frac{15\pi}{180}, \frac{25\pi}{180}, \cdots, \frac{95\pi}{180} \right\}.\] 
	
	\begin{figure}[t]
		\centering
		\includegraphics[width=0.85\textwidth]{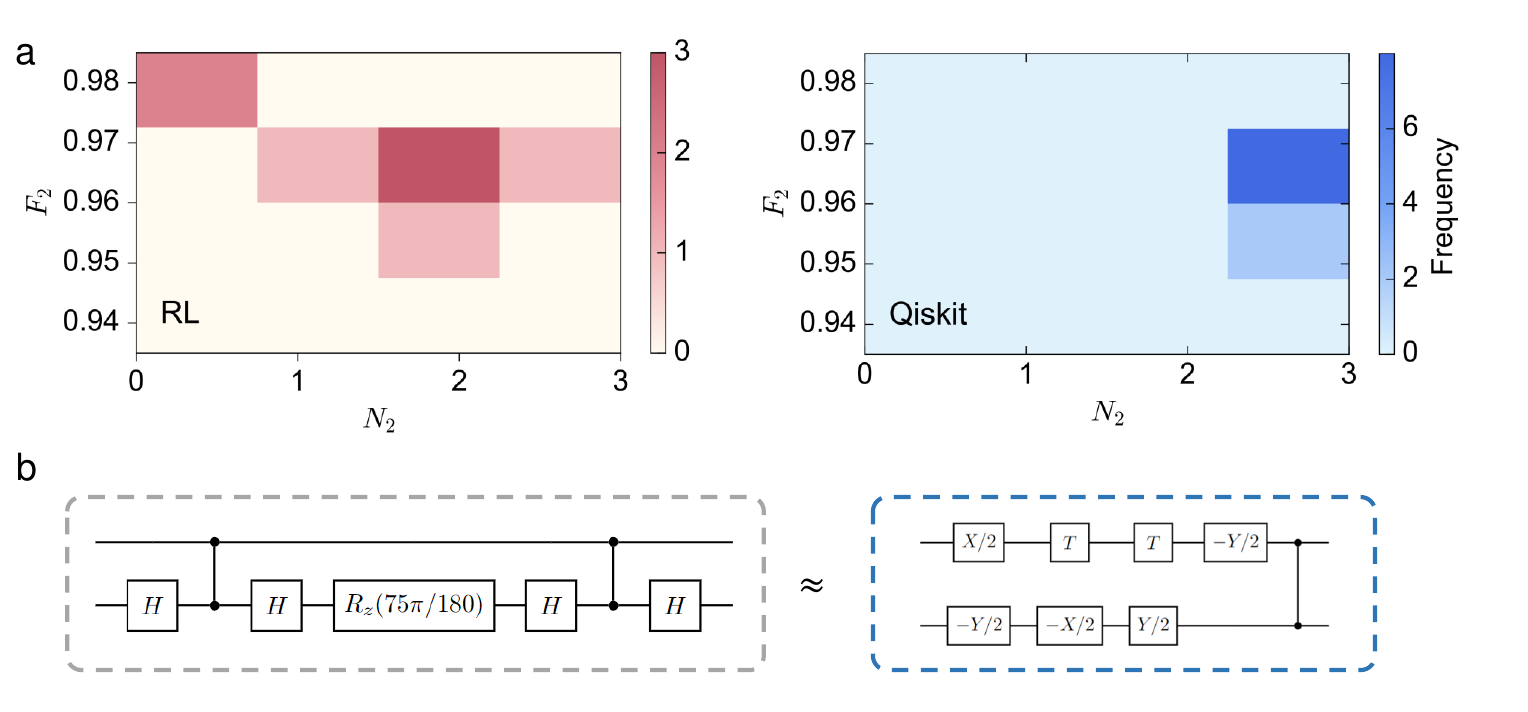}
		\caption{\small{\textbf{Compilation of $\RZZ(\gamma)$ gates.} (a) Frequency under variations of experimental fidelity $F_2$ and two-qubit gate count $N_2$ for circuits by RL-based compiler (left panel) and Qiskit compiler (right panel). (b) Left side: target unitary; right side: gate sequence by RL-based compiler with $F_1=0.9914$. }}
		\label{fig:SM_Rzz_noisy}
	\end{figure}
	
	The experimental results are shown in Fig.~\ref{fig:SM_Rzz_noisy}(a). For 10 target gates, the RL-based compiler reaches the accuracy with ${F_1}=0.9967 \pm 0.0029$ and $F_2=0.9615 \pm 0.0153$ on average, where the corresponding two-qubit gate count is $N_2=1.6$ on average. The average inference time of AQ* search is around $480$ seconds. For the Qiskit compiler, the accuracy is $F_1= 1$ and $F_2 = 0.9633 \pm 0.0036$, respectively, and the number of two-qubit gates is $N_2=3$. Hence, except for one target unitary {$\RZZ(25\pi/180)$}, the RL-based compiler requires fewer two-qubit gates than those required by the Qiskit compiler, while the achieved compiling accuracy is comparable. We show the gate decomposition of $\RZZ(75\pi/180)$ returned by the RL-based compiler in Fig.~\ref{fig:SM_Rzz_noisy}(b). Compared with the standard solution requiring three two-qubit gates, the RL-based compiler is more resource-efficient and fidelity-warranted. 
	
	\begin{figure*}[b]
	\centering
	\includegraphics[width=0.9\textwidth]{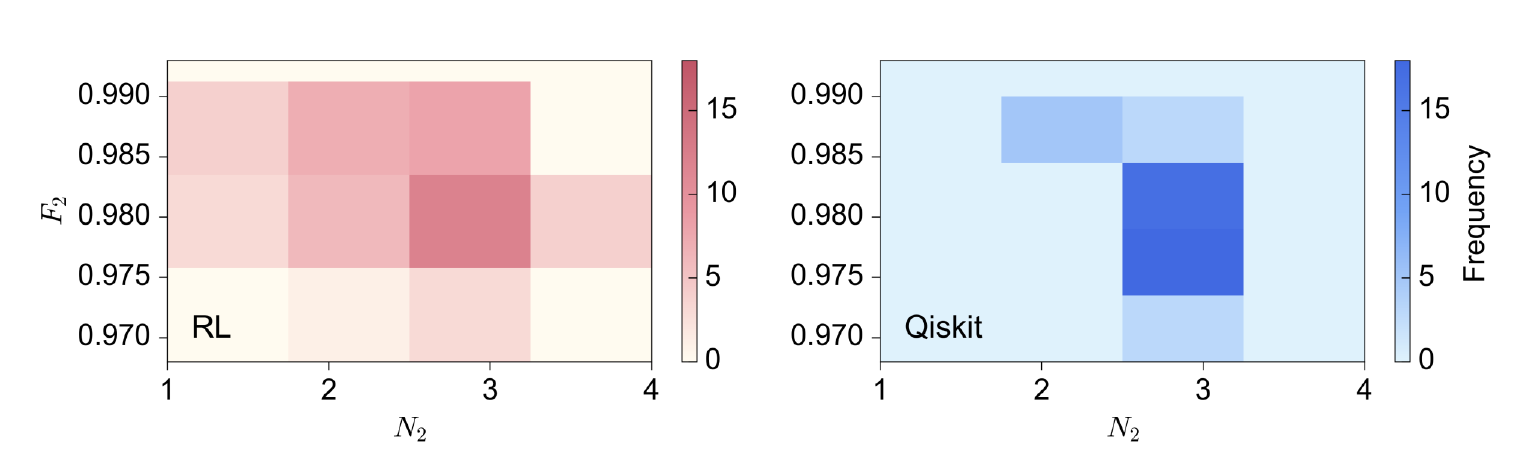}
	\caption{\small{\textbf{Compilation of special targets for robust verification.} Frequency under variations of experimental fidelity $F_2$ and two-qubit gate count $N_2$ for circuits by RL-based compiler (left panel) and Qiskit compiler (right panel).}}
	\label{fig:SM_layer_40}
    \end{figure*}

	\medskip
	\noindent\textbf{Task 4: More special targets for robust verification}. In Task~1 where the targets are composed of gates randomly sampled from $\mathcal{G}^{(2,1)}$, the RL-based compiler demonstrates a slight benefit over the Qiskit compiler in terms of the exploited two-qubit gates. In Task~2 where the targets are randomly sampled from SU(4), our proposal shows a slight disadvantage. Hence, a question that naturally arises is whether the performance of the RL-based compiler can be enhanced even further by reducing the complexity of target unitaries. Here we answer this question by generating a new test dataset containing $50$ target unitaries, i.e., the construction rule is the same as the one used in the main text, while the only difference is randomly sampling $40$ single-qubit gates from $\mathcal{G}^{(2,1)}$ rather than $80$ single-qubit gates. 
	
	The experimental results are exhibited in Fig.~\ref{fig:SM_layer_40}. On average, the RL-based compiler uses $N_2=2.38$ CZ gates to accomplish the compiling task with fidelity $F_2=0.9824\pm 0.0058$, while the Qiskit compiler uses $N_2=2.9$ CZ gates with fidelity $F_2=0.9789\pm 0.0045$. The results indicate that a well-trained DQN with more training data and deeper loop number $L$ contributes to enhancing the robustness in compiling more complicated circuits. 

    \subsection{Summary of results for two-qubit system}\label{append:subsec:sum_two_exp}

    We summarize the results of compilation for the two-qubit system in Tab.~\ref{tab:table4}.
	
	\begin{table}[t]
		\caption{\label{tab:table4}
			\small{\textbf{Performance of RL-based compiler and Qiskit compiler for the two-qubit system.} $F_2$(XEB) and $F_2$(QPF) are the outcomes from XEB measurement and QPT measurement, respectively.}}
		\begin{ruledtabular}
			\begin{tabular}{c|ccccc|ccccc}
				& \multicolumn{5}{c|}{Qiskit}&\multicolumn{5}{c}{RL}\\
				\hline
				Target & $N_2$& $N_1$ & $F_1$&$F_2$(XEB)
				& $F_2$(QPT) & $N_2$& $N_1$ & $F_1$&$F_2$(XEB)&$F_2$(QPT) \\
				\hline
				SWAP&3&12&1&0.951&0.985 &3&6&1&0.969&0.986\\
				$\RZZ({\pi}/{2})$ & 2 & 8 & 1 & 0.973 & 0.981 &1 & 4 & 1 &0.982&0.988 \\
				40 basis gates$\rightarrow$ U & 2.9 & 15.6 &0.9999997&0.956 & 0.979& 2.38 & 13.1& 0.999995& 0.967 & 0.982 \\
				80 basis gates$\rightarrow$ U & 2.957 & 15.84 &0.9999997&0.954 & 0.979 & 2.80 & 11.8&  0.999991& 0.966&0.984 \\
				Random SU(4) & 3 & 16.0 &0.9999997 &0.953 & 0.978 & 2.97 & 14.2& 0.9911 & 0.933 &0.963\\
			\end{tabular}
		\end{ruledtabular}
	\end{table}
	
\vspace{10mm}
\section{EXTENDED DISCUSSION AND RESULTS FOR THREE-QUBIT SYSTEM}\label{appendix:three-qubit-res}
\vspace{5mm}

In this section, we describe the construction rule of the training data and test data, and the implementation details of the RL-based compiler. A summary of the results of compilation for the three-qubit system is also presented.
	
	\subsection{Construction of training data and test data}\label{appendix:subD1:data-generation}
	
	\noindent\textbf{Data generation}. We exploit four basis sets for different three-qubit compiling tasks. The first basis set is $\mathcal{G}^{(3,1)}= \{ \pm \X/2,\pm \Y/2, \T, \T^{\dagger}, \CZ \}$, where the superscript $(3,1)$ represents the three-qubit case under the first basis set and the connectivity constraint is $\rm Q_1\mbox{-}Q_2\mbox{-}Q_3$. Thus the CZ gate can only be implemented on neighboring qubits.  Accordingly, the generalized basis set, or equivalently the action space, yields
	\begin{equation}
		\begin{aligned}
			\tilde{\mathcal{G}}^{(3, 1)}\equiv\mathcal{A}^{(3,1)} = \Big \{& \RX(\frac{\pi}{2})\otimes \mathbb{I}_4, \mathbb{I}_2\otimes \RX(\frac{\pi}{2})\otimes \mathbb{I}_2, \mathbb{I}_4 \otimes \RX(\frac{\pi}{2}), \RX(-\frac{\pi}{2})\otimes \mathbb{I}_4, \mathbb{I}_2\otimes \RX(-\frac{\pi}{2})\otimes \mathbb{I}_2, \mathbb{I}_4\otimes \RX(-\frac{\pi}{2}),    \\
			& \RY(\frac{\pi}{2})\otimes \mathbb{I}_4, \mathbb{I}_2\otimes \RY(\frac{\pi}{2})\otimes \mathbb{I}_2, \mathbb{I}_4\otimes \RY(\frac{\pi}{2}), \RY(-\frac{\pi}{2})\otimes \mathbb{I}_4, \mathbb{I}_2\otimes \RY(-\frac{\pi}{2})\otimes \mathbb{I}_2, \mathbb{I}_4\otimes \RY(-\frac{\pi}{2}), \\ 
			& T\otimes \mathbb{I}_4, \mathbb{I}_2\otimes T\otimes \mathbb{I}_2, \mathbb{I}_4\otimes T, T^{\dagger}\otimes \mathbb{I}_4, \mathbb{I}_2\otimes T^{\dagger}\otimes \mathbb{I}_2, \mathbb{I}_4\otimes T^{\dagger}, \CZ\otimes \mathbb{I}_2, \mathbb{I}_2\otimes \CZ \Big \},
		\end{aligned}
	\end{equation}
	with the cardinal number $|\mathcal{A}^{(3,1)}|=20$. 
	
The second basis set is formed by HRC gate set \cite{harrow2002efficient} with $\mathcal{G}^{(3, 2)}=\{V_1, V_1^{\dagger}, V_2, V_2^{\dagger}, V_3, V_3^{\dagger}, \CZ \}$, where 
	\begin{equation}
		\begin{aligned} V_1 & =\frac{1}{\sqrt{5}}\left(\begin{array}{cc}1 & 2 i \\ 2 i & 1\end{array}\right), 
			V_2=\frac{1}{\sqrt{5}}\left(\begin{array}{cc}1 & 2 \\ -2 & 1\end{array}\right), 
			V_3 & =\frac{1}{\sqrt{5}}\left(\begin{array}{cc}1+2 i & 0 \\ 0 & 1-2 i\end{array}\right) .\end{aligned}
	\end{equation}
	Considering the same connectivity constraint, the corresponding action space yields 
	\begin{equation}
		\begin{aligned}
			\mathcal{A}^{(3,2)}=\Big\{& V_1\otimes \mathbb{I}_4, \mathbb{I}_2\otimes V_1\otimes \mathbb{I}_2, \mathbb{I}_4 \otimes V_1, V_1^{\dagger}\otimes \mathbb{I}_4, \mathbb{I}_2\otimes V_1^{\dagger}\otimes \mathbb{I}_2, \mathbb{I}_4\otimes V_1^{\dagger},  V_2\otimes \mathbb{I}_4, \mathbb{I}_2\otimes V_2\otimes \mathbb{I}_2, \mathbb{I}_4\otimes V_2,       V_2^{\dagger}\otimes \mathbb{I}_4, \\
			& \mathbb{I}_2\otimes V_2^{\dagger}\otimes \mathbb{I}_2, \mathbb{I}_4\otimes V_2^{\dagger}, V_3\otimes \mathbb{I}_4, \mathbb{I}_2\otimes V_3\otimes \mathbb{I}_2, \mathbb{I}_4\otimes V_3, V_3^{\dagger}\otimes \mathbb{I}_4, \mathbb{I}_2\otimes V_3^{\dagger}\otimes \mathbb{I}_2, \mathbb{I}_4\otimes V_3^{\dagger}, \CZ\otimes \mathbb{I}_2, \mathbb{I}_2\otimes \CZ \Big\},  
		\end{aligned}
	\end{equation}
	with the cardinality $|\mathcal{A}^{(3,2)}|=20$.
	
For the third basis set, we consider the ideal setting by removing the topological constraint of $ \mathcal{G}^{(3,2)}$, with $\mathcal{G}^{(3, 3)}=\mathcal{G}^{(3, 2)}=\{V_1, V_1^{\dagger}, V_2, V_2^{\dagger}, V_3, V_3^{\dagger}, H, \CZ \}$. Accordingly, the action space with respect to the third basis set yields
	\begin{equation}
		\begin{aligned}
			\mathcal{A}^{(3,3)}=\Big \{& V_1\otimes \mathbb{I}_4, \mathbb{I}_2\otimes V_1\otimes \mathbb{I}_2, \mathbb{I}_4 \otimes V_1, V_1^{\dagger}\otimes \mathbb{I}_4, \mathbb{I}_2\otimes V_1^{\dagger}\otimes \mathbb{I}_2, \mathbb{I}_4\otimes V_1^{\dagger}, V_2\otimes \mathbb{I}_4, \mathbb{I}_2\otimes V_2\otimes \mathbb{I}_2, \mathbb{I}_4\otimes V_2,     \\
			& V_2^{\dagger}\otimes \mathbb{I}_4, \mathbb{I}_2\otimes V_2^{\dagger}\otimes \mathbb{I}_2, \mathbb{I}_4\otimes V_2^{\dagger},  V_3\otimes \mathbb{I}_4, \mathbb{I}_2\otimes V_3\otimes \mathbb{I}_2, \mathbb{I}_4\otimes V_3, V_3^{\dagger}\otimes \mathbb{I}_4, \mathbb{I}_2\otimes V_3^{\dagger}\otimes \mathbb{I}_2, \mathbb{I}_4\otimes V_3^{\dagger}, \\ 
			& \text{H}\otimes \mathbb{I}_4, \mathbb{I}_2\otimes \text{H}\otimes \mathbb{I}_2, \mathbb{I}_4\otimes \text{H}, \CZ\otimes \mathbb{I}_2, \mathbb{I}_2\otimes \CZ, \ket{0}\bra{0}\otimes \mathbb{I}_4 + \ket{1}\bra{1}\otimes \mathbb{I}_2\otimes \Z \Big\},
		\end{aligned}
	\end{equation}
	with the cardinality $|\mathcal{A}^{(3,3)}|=24$. 
	
The fourth gate set is formed by the single-qubit rotation gate  with $\mathcal{G}^{(3,4)}= \{ \pm \X/128,\pm \Y/128, \pm \Z/128 \}$. Accordingly, the action space for the fourth basis gate set yields
	\begin{equation}
		\begin{aligned}
			\mathcal{A}^{(3,4)}=\Big \{& \RX(\frac{\pi}{128})\otimes \mathbb{I}_4, \mathbb{I}_2\otimes \RX(\frac{\pi}{128})\otimes \mathbb{I}_2, \mathbb{I}_4 \otimes \RX(\frac{\pi}{128}), \RX(-\frac{\pi}{128})\otimes \mathbb{I}_4, \mathbb{I}_2\otimes \RX(-\frac{\pi}{128})\otimes \mathbb{I}_2, \mathbb{I}_4\otimes \RX(-\frac{\pi}{128}),    \\
			& \RY(\frac{\pi}{128})\otimes \mathbb{I}_4, \mathbb{I}_2\otimes \RY(\frac{\pi}{128})\otimes \mathbb{I}_2, \mathbb{I}_4\otimes \RY(\frac{\pi}{128}), \RY(-\frac{\pi}{128})\otimes \mathbb{I}_4, \mathbb{I}_2\otimes \RY(-\frac{\pi}{128})\otimes \mathbb{I}_2, \mathbb{I}_4\otimes \RY(-\frac{\pi}{128}), \\ 
			& \RZ(\frac{\pi}{128})\otimes \mathbb{I}_4, \mathbb{I}_2\otimes \RZ(\frac{\pi}{128})\otimes \mathbb{I}_2, \mathbb{I}_4\otimes \RZ(\frac{\pi}{128}), \RZ(-\frac{\pi}{128})\otimes \mathbb{I}_4, \mathbb{I}_2\otimes \RZ(-\frac{\pi}{128})\otimes \mathbb{I}_2, \mathbb{I}_4\otimes \RZ(-\frac{\pi}{128}) \Big \}
		\end{aligned}
	\end{equation}
	with the cardinality $|\mathcal{A}^{(3,4)}|=18$.
	
\begin{figure*}[t]
		\centering
		\includegraphics[width=0.89\textwidth]{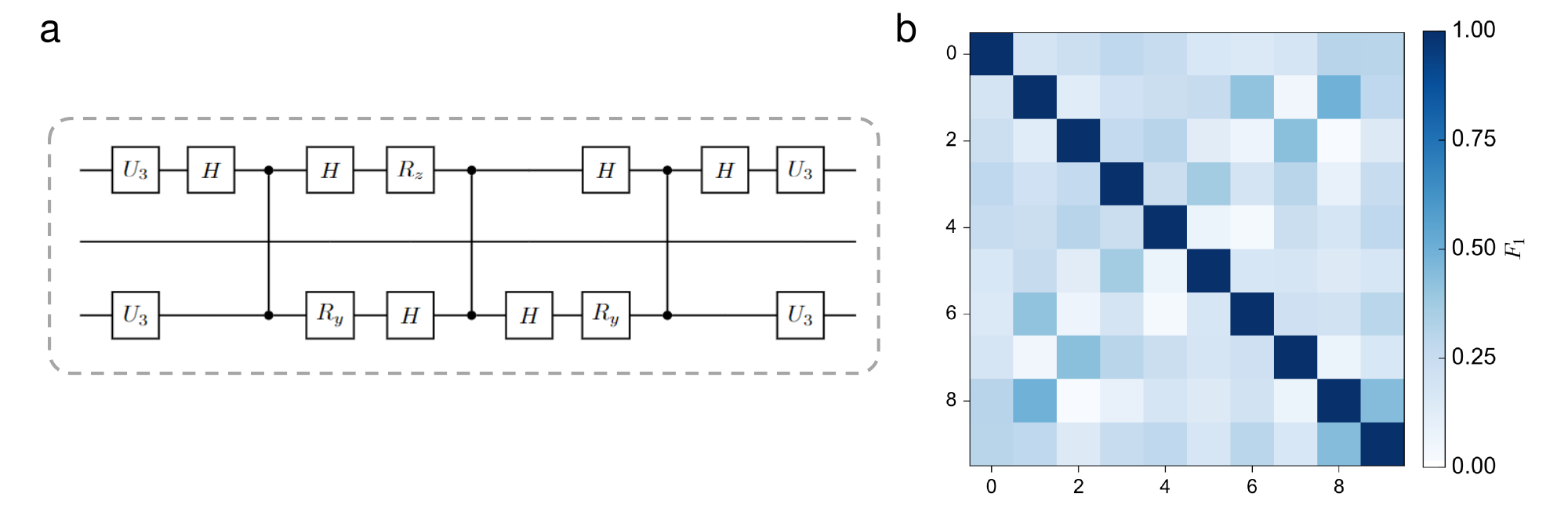}
		\caption{\small{\textbf{Three-qubit compiling targets generation.} (a) Template circuit for compiling target generation. (b) The heatmap of the similarity between 10 target unitaries employed in the three-qubit task. The $F_1$ metric quantifies the similarity.
		}}
		\label{fig:SM_three_qubits_targets_generation}
	\end{figure*}

    \medskip	
	\noindent\textbf{Compiling targets generation}. The construction rule of three-qubit targets is similar to that of the two-qubit case. As shown in Fig.~\ref{fig:SM_three_qubits_targets_generation}(a), we consider the two-qubit gates operated on $\rm Q_1$ and  $\rm Q_3$. As such, we collect the three-qubit targets by fixing three CNOT gates operated on $\rm Q_1$ and  $\rm Q_3$, and operating those single-qubit gates on $\rm Q_1$ and  $\rm Q_3$. Particularly, 20 single-qubit gates uniformly sampled from $\tilde{\mathcal{G}}^{(3, 1)}$ are employed to form the single-qubit gates in the circuit. Note that due to the topological constraint $\rm Q_1\mbox{-}Q_2\mbox{-}Q_3$, this task belongs to the three-qubit gate compiling. Fig.~\ref{fig:SM_three_qubits_targets_generation}(b) exhibits the dissimilarity of the employed 10 unitaries used in the three-qubit gate compiling tasks. 
	
	\subsection{Implementation of RL-based compiler}\label{appendix:subD2:implentation-details}
	
	\noindent\textbf{Classical Device}. Both the precompilation and inference processes related to the classical part are implemented on the Nvidia Tesla V100 GPU.  
	
	\medskip
	\noindent\textbf{Architecture of DQNs.} The implementations of DQNs in three-qubit gate compiling tasks are the same as those used in two-qubit tasks except for the model size. The visualization of the employed DQN is shown in Fig.~\ref{fig:dqn_structure}. Specifically, it consists of an input layer, two hidden layers, and six residual blocks,  followed by $|\mathcal{A}^{(3, t)}|$ output neurons with $t\in\{1, 2, 3, 4\}$. The input layer is of size $64$. The first two hidden layers are of sizes 8000 and 3000, and each residual block is composed of two fully connected layers with 3000 hidden neurons. The leaky ReLU activation function and batch normalization are applied to all hidden layers.  
	
	\medskip
	\noindent\textbf{Hyper-parameters}. 
	We use Adam optimizer to optimize DQN, where the learning rate is $\eta =10^{-3}$ without the weight decay \cite{kingma2017adam}. The loss threshold $\delta$ in Alg.~\ref{alg:train_DQN} is set as $0.02$. The epoch $T$ depends on $l$, i.e., $T=100\cdot l$. In the training process, the optimized DQN models at each loop $l$ under each action space $\mathcal{A}^{(3,t)}$, denoted by $\{Q^{(t, {l})}_{\bm{\theta}}\}$, are stored to conduct multi-DQN collaborative AQ* search. In the inference stage, we execute the multi-DQN collaborative AQ* search guided by DQN model $\{Q^{(t, {l})}_{\bm{\theta}}\}$. And the maximum AQ* search depth is set as 500 and the inference time is restricted to be less than 20 minutes.
	
	\medskip
	\noindent\textbf{The total variation distance (TVD).}{\label{TVD}}
	We exploit the total variation distance to measure the distance between two distributions in the main text. To be concrete, the total variation distance (TVD) is a standard metric to measure the statistical distance between two probability distributions:
	\begin{equation}
		d_{TV}(P_{e},P_{i}) = \frac{1}{2} \sum_{x} |P_{e}(x)-P_{i}(x)|,
	\end{equation}
	where in this context, $P_{i}(x)$ and $P_{e}(x)$ denote the distributions of the ideal and experimental quantum states in the measure of computational basis, respectively.
	
    \subsection{Summary of results for three-qubit system}\label{append:subsec:sum_three_exp}

    We summarize the results of compilation for the three-qubit system in Tab.~\ref{tab:table5}. 

	\begin{table}[t]
		\caption{\label{tab:table5}
			\small{\textbf{Performance of Qiskit, RL-based, and VRL-based compilers for three-qubit system.}}}
		\begin{ruledtabular}
			\begin{tabular}{c|cccc|cccc|cccc}
				& \multicolumn{4}{c|}{Qiskit}&\multicolumn{4}{c
					|}{RL}&\multicolumn{4}{c}{VRL}\\
				\hline
				Target & $N_2$& $N_1$ & $F_1$
				& $F_2$(QPT) & $N_2$& $N_1$ & $F_1$&$F_2$(QPT) & $N_2$& $N_1$ & $F_1$
				& $F_2$(QPT) \\
				\hline
				QFT&15&40&1&0.739&7&28 &0.94&0.834&7&28&1-10$^{-7}$&0.932\\
				Topologically constrained&35.7&65.3&1&0.626 &6.36 & 20.2&1&0.934&$-$&$-$&$-$&$-$\\
			\end{tabular}
			
		\end{ruledtabular}
	\end{table}
	
\vspace{10mm}
\section{VARIATIONAL RL-BASED QUANTUM COMPILER}\label{sm_last_sebsec_vqa}
\vspace{5mm}

The variational principle is widely used in the studies of many-body physics and chemistry~\cite{mclachlan1964variational}. The key concept of the variational methods is considering trial states from a physically motivated small subset of the exponentially large Hilbert space, which effectively avoids the exponential space problem. Examples include variational quantum algorithms~\cite{cerezo2021variational,bharti2022noisy,abbas2020power,huang2021power,du2022demystify,cerezo2022challenges,tian2023recent} and tensor networks~\cite{Huang2021PRL, feng2020Contracting, Yong2021Closing, Thorsten2023Simulating, Feng2022Solving}.  

We have combined the variational methods with the RL-based compiler using a toolkit to optimize the single-qubit gate parameters classically. Below we show that such variational RL-based (VRL-based) compiler can reach a higher circuit fidelity, albeit with a trade-off of the increased inference time. 
 	
	\medskip
	\noindent\textbf{Implementation}. The implementation of the VRL-based compiler consists of two steps. In the first step, the optimized RL-based compiler described in Alg.~\ref{alg:train_DQN} is utilized to search for a gate sequence $\{\tilde{G}_i\}_{i=1}^{N}\in\tilde{\mathcal{G}}^N$ that approximates the target $M$-qubit unitary $\hat{U}$. The estimated unitary is denoted by $\hat{V}=\prod_{i=N}^1 \tilde{G}_i\in \mathbb{C}^{2^M\times 2^M}$. In the second step, the searched gate sequence $\{\tilde{G}_i\}_{i=1}^{N}$ is transformed into a variational ansatz, i.e., $\hat{V}_{\bm{\gamma}}$, by parametrizing the quantum gates in $\{\tilde{G}_i\}_{i=1}^{N}$. Note that the way of parameterization is flexible and a possible approach is detailed below. Once the parameterization is completed, the trainable parameters $\bm{\gamma}$ are updated to minimize the loss function that measures the discrepancy between the estimated unitary $\hat{V}_{\bm{\gamma}}$ and the target unitary $\hat{U}$. The explicit form of the loss function is 
	\begin{eqnarray}\label{eqn:loss_vqa}
		\min_{\bm{\gamma}}\mathcal{L}(\hat{V}_{\bm{\gamma}}, \hat{U}) = 2^M-\Tr\left( \hat{U}^{\dagger} \cdot \hat{V}_{\bm{\gamma}} \right).
	\end{eqnarray}

	\medskip
	\noindent\textbf{Construction of the variational ansatz $\hat{V}_{\bm{\gamma}}$}. Given the searched gate sequence $\{\tilde{G}_i\}_{i=1}^{N}$, the construction rule of the variational ansatz $\hat{V}_{\bm{\gamma}}$ is heuristic. All the single- and two-qubit gates can be parameterized to form a variational ansatz. Practically,  we only need to parameterize the single-qubit gates in $\{\tilde{G}_i\}_{i=1}^{N}$ for higher circuit fidelity and keep the two-qubit gate parameters fixed.
		
\begin{figure*}[t]
		\centering
		\includegraphics[width=0.99\textwidth]{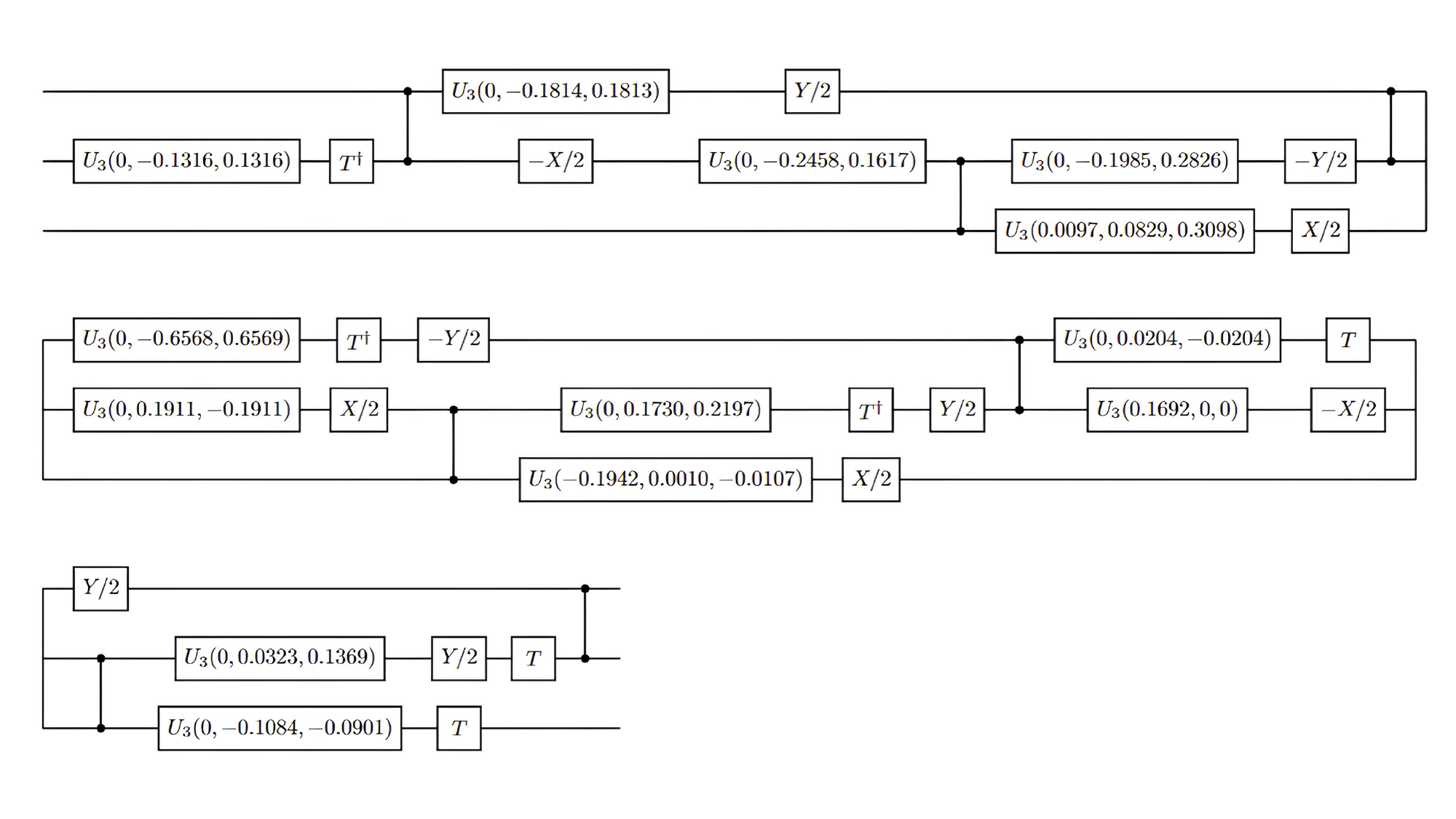}
		\caption{\small{\textbf{Circuit of three-qubit QFT compiled by VRL-based compiler.} Without the $\rm U_3$ gates, the circuit is returned by the RL-based compiler.
		}}
		\label{fig:exp_vqa_two_qubits}
	\end{figure*}

	\medskip
	\noindent\textbf{Results}. We have seen in the main text that for the three-qubit QFT operation, the fidelity of the RL compiled circuit is $F_1=0.94$. In combining the variational method, the settings of the RL quantum compiler are identical to those introduced in the main text. The Adam optimizer~\cite{kingma2017adam} with the initial learning rate $0.01$ is employed to minimize the loss function in Eq.~(\ref{eqn:loss_vqa}). The learning steps are fixed to be $500$ and the process is repeated five times to suppress the effects of randomness. Figure~\ref{fig:exp_vqa_two_qubits} illustrates the gate sequence returned by the VRL-based compiler. In Tab.~\ref{tab:table5}, we find that the fidelity of the circuit by the VRL-based compiler is $F_1=1-10^{-7}$ and the corresponding experimental fidelity is $F_2=0.932$. The inference time increases from about 800 seconds to about 900 seconds.
	
\end{document}